\documentclass[3p,twocolumn]{elsarticle}
\makeatletter
\def\set@curr@file#1{%
  \begingroup
    \escapechar\m@ne
    \xdef\@curr@file{\expandafter\string\csname #1\endcsname}%
  \endgroup
}
\def\quote@name#1{"\quote@@name#1\@gobble""}
\def\quote@@name#1"{#1\quote@@name}
\def\unquote@name#1{\quote@@name#1\@gobble"}
\makeatother

\usepackage{verbatim}
\usepackage{graphics} % for pdf, bitmapped graphics files
\usepackage{epsfig} % for postscript graphics files
\usepackage{epstopdf}
\usepackage{subfigure}
\usepackage{epstopdf}
\usepackage{balance}
\usepackage{comment}
\usepackage{booktabs}
\usepackage{amsmath}
\usepackage{amssymb}
\usepackage{amsfonts}
\usepackage{bm}
\usepackage{colortbl}
\usepackage{tabularx}
\usepackage{color}
\usepackage{epsfig}
\usepackage{xspace}
\usepackage{graphicx}    % For importing graphics
\usepackage{algorithm}
\usepackage{algpseudocode}
\usepackage{placeins}
\usepackage{multirow}
\usepackage{caption}
\usepackage{enumitem}
\usepackage{leading}
\usepackage{xcolor}

\begin{document}

\begin{frontmatter}

\title{A Multi-view CNN-based Acoustic Classification System for Automatic Animal Species Identification}

  \author[1]{Weitao Xu}
\ead{weitaoxu@cityu.edu.hk}

 \author[2]{Xiang Zhang}
 \ead{xiang.zhang3@student.unsw.edu.au}

\author[2]{Lina Yao}
 \ead{lina.yao@unsw.edu.au}

 \author[2]{Wanli Xue}
 \ead{w.xue@unsw.edu.au}

 \author[3]{Bo Wei}
 \ead{bo.wei@northumbria.ac.uk}

\address[1]{Department of Computer Science, City University of Hong Kong, Hong Kong}
\address[2]{School of Computer Science and Engineering, University of New South Wales, Australia}
\address[3]{Department of Computer and Information Sciences, Northumbria University, UK}

\begin{abstract}
Automatic identification of animal species by their vocalization is an important and challenging task. Although many kinds of audio monitoring system have been proposed in the literature, they suffer from several disadvantages such as non-trivial feature selection, accuracy degradation because of environmental noise or intensive local computation. In this paper, we propose a deep learning based acoustic classification framework for Wireless Acoustic Sensor Network (WASN). The proposed framework is based on cloud architecture which relaxes the computational burden on the wireless sensor node. To improve the recognition accuracy, we design a multi-view Convolution Neural Network (CNN) to extract the short-, middle-, and long-term dependencies in parallel. The evaluation on two real datasets shows that the proposed architecture can achieve high accuracy and outperforms traditional classification systems significantly when the environmental noise dominate the audio signal (low SNR). Moreover, we implement and deploy the proposed system on a testbed and analyse the system performance in real-world environments. Both simulation and real-world evaluation demonstrate the accuracy and robustness of the proposed acoustic classification system in distinguishing species of animals.
\end{abstract}

\begin{keyword}
Wireless acoustic sensor network \sep Animal identification \sep Deep learning \sep CNN
\end{keyword}
\end{frontmatter}

\section{Introduction}
\label{sec:intro}
Wireless Acoustic Sensor Network (WASN) based animal monitoring is of great importance for biologists to monitor real-time wildlife behavior for long periods and under variable weather/climate conditions. The acquired animal voice can provide valuable information for researchers, such as the density and diversity of different species of animals~\cite{hao2013monitoring,luque2016evaluation,akyildiz2005underwater}. For example, Hu et al. proposed a WASN application to census the populations of native frogs and the invasive introduced species (Cane Toad) in Australia~\cite{hu2009design}. 
There are also several important commercial applications of acoustic animal detection. For instance, America imports billions of dollars of timber from Aisa every year. However, the inadvertent introduction of the Asian Longhorn Beetle has cost USA government millions of dollars to eradicate the Beetle population~\cite{nowak2001potential}. Therefore, a wireless monitoring system is imperative to detect the distribution of these insects.

There are a large volume of audio monitoring systems in the literature~\cite{hu2009design,anderson1996template,kogan1998automated,fagerlund2007bird,guo2003content,huang2009frog,acevedo2009automated,banerjee2014partial,dutta2013energy,diaz2012use}. In the early stage, biologists have traditionally deployed audio recording systems over the natural environment where their research projects were  developed~\cite{anderson1996template,kogan1998automated}.
However this procedure requires human presence in the area of interest at certain moments. In recent years, with the development of WSN, some researchers have proposed remotely accessible systems in order to minimize the impact of the presence of human beings in the habitat of interest~\cite{hu2009design,banerjee2014partial,dutta2013energy,diaz2012use}. 

Despite much effort in this area, previous studies suffer from several disadvantages. First, traditional methods usually first extract a number of appropriate features and then employ classic machine learning methods such as Support Vector Machine (SVM) or K-Nearest Neighbours (KNN) to detect the species of the animals. Features, such as statistical features through statistical anlaysis (e.g., variance, mean, median), Fast Fourier Transmission (FFT) spectrum, spectrograms, Wigner-Ville distribution (WVD), Mel-frequecy cepstrum coefficient (MFCC) and wavelets have been broadly used. However, extracting robust features to recognizing noisy field recordings is non-trivial. While these features may work well for one , it is not clear whether they generalize to other species. The specific features for one application do not necessarily generalize to others. Moreover, a significant number of calibrations are required for both manually feature extraction and the classification algorithms. This is because the performance of the traditional classifiers such as SVM and KNN~\cite{fagerlund2007bird,huang2009frog,acevedo2009automated} highly depends on the quality of the extracted features. However, handcrafting features relies on a significant amount of domain and engineering knowledge to translate insights into algorithmic methods. Additionally, manual selection of good features is slow and costly in effort. Therefore, these approaches lack scalability for new applications. Deep learning technologies can solve these problems by using deep architectures to learn feature hierarchies. The features that are higher up in deep hierarchies are formed by the composition of features on lower levels. These multi-level representations allow a deep architecture to learn the complex functions that map the input (such as digital audio) to output (e.g. classes), without the need of dependence on manual handcrafted features.

Secondly, these approaches suffer from accuracy degradation in real-world applications because of the impact of environmental noise. The voice recorded from field usually contains much noise which poses a big challenge to real deployment of such system. To address this problem, Wei et al.~\cite{wei2013real} proposed an \textit{in-situ} animal classification system by applying sparse representation-based classification (SRC). SRC uses $\ell_1$-optimization to make animal voice recognition robust to environmental noise. However, it is known that $\ell_1$-optimization is computationally expensive~\cite{shen2014face,xu2016sensor}, which limits the application of their system in resource-limited sensor nodes. Additionally, in order to make SRC achieve high accuracy, a large amount of training data is required. This means a wireless sensor node can only store a limited number of training classes because of the limited storage.

Recently, deep learning has emerged as a powerful tool to solve various recognition tasks such as face recognition~\cite{sun2014deep}, human speech recognition~\cite{hinton2012deep,graves2013speech} and natural language processing~\cite{collobert2008unified}. The application of deep learning in audio signal is not new; however, most previous studies focus on human speech analysis to obtain context information~\cite{hinton2012deep,graves2013speech,wang2016attention}. Limited efforts have been devoted to applying deep learning in WASN to classify different species of animals. To bridge this gap, we aim to design and implement a acoustic classification framework for WASN by employing deep learning techniques. Convolutional Neural Network (CNN), as a typical deep learning algorithm, has been widely used in high-level representative feature learning. In detail, CNN is enabled to capture the local spatial coherence from the input data. In our case, the spatial information refers to the spectral amplitude of the audio signal. However, one drawback of the standard CNN structure is that the filter length of the convolution operation is fixed. As a result, the convolutional filter can only discover the spatial features with the fixed filter range. For example, CNN may explore the short-term feature but fail to capture the middle- and long-term features. In this paper, we propose a multi-view CNN framework which contains three convolution operation with three different filter length in parallel in order to extract the short-, middle-, and long-term information at the same time. We conduct extensive experiments to evaluate the system on real datasets. More importantly, we implement the proposed framework on a testbed and conduct a case study to analyse the system performance in real environment. To the best of our knowledge, this is the first work that designs and implements a deep learning based acoustic classification system for WASN.  

The main contributions of this paper are threefold:
\begin{itemize}
	\item We design a deep learning-based acoustic classification framework for WASN, which adopts a multi-view convolution neural network in  order  to  automatically  learn  the  latent  and  high-level  features from  short-, middle- and long-term audio signals in parallel.
	\item We conduct extensive evaluation on two real dataset (Forg dataset and Cricket dataset) to demonstrate the classification accuracy and robustness of the proposed framework to environmental noise. Evaluation results show that the proposed system can achieve high recognition accuracy and outperform traditional methods significantly especially in low SNR scenarios.
	\item We implement the proposed system on a testbed and conduct a case study to evaluate the performance in real world environments. The case study demonstrate that the proposed framework can achieve high accuracy in real applications.
\end{itemize}
The rest of this paper is organized as follows. Section~\ref{sec:relatedwork} introduces related work. Then, we describe system architecture in Section~\ref{sec:system} and evaluate the system performance in Section~\ref{sec:evaluation}. We implement the system on a testbed and conduct user study to evaluate the system in Section~\ref{sec:testbed}. Finally, Section~\ref{sec:conclusion} concludes the paper.
\begin{figure*}[!ht]
    \centering
         \includegraphics[width=6.5in]{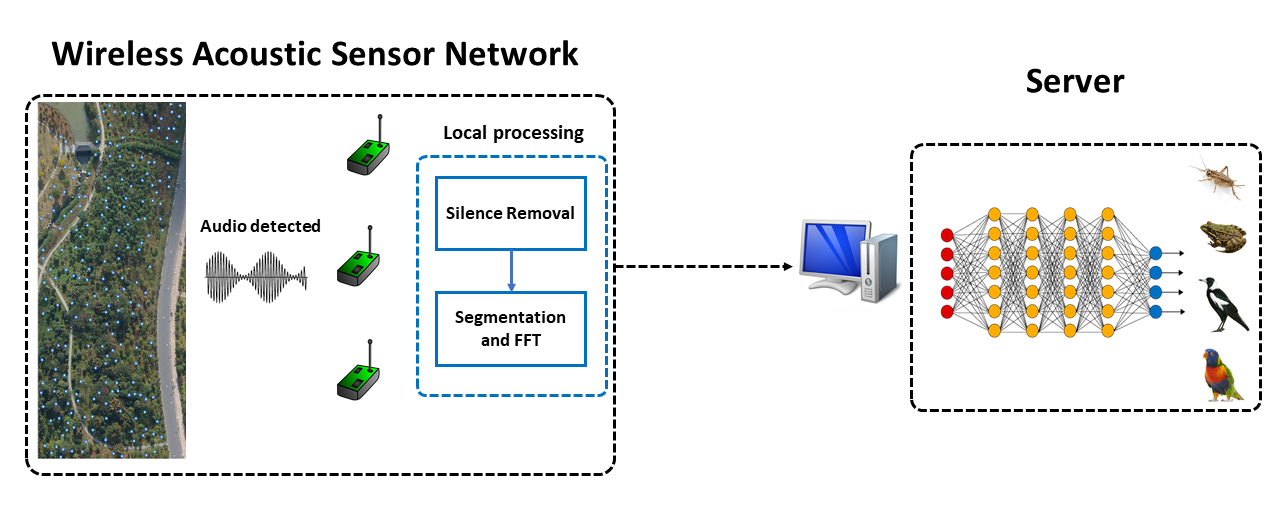}
    \caption{System Overview.}
   \label{fig:systemoverview}
\end{figure*}
\section{Related Work}
\label{sec:relatedwork}
Animal voice classification has been extensively studied in the literature. At the highest level, most work extract sets of features from the data, and use these features as inputs for standard classification algorithms such as SVM, KNN, decision tree, or Bayesian classifier. Previous studies have involved a wide range of species which include farm animals~\cite{jahns1997sound}, bats~\cite{preatoni2005identifying}, birds~\cite{fagerlund2007bird,acevedo2009automated,hodon2015monitoring}, pests~\cite{eliopoulos2016estimation}, insects~\cite{ntalampiras2019automatic} and anurans~\cite{vaca2010using}. The works of Anderson et al.~\cite{anderson1996template} and Kogan et al.~\cite{kogan1998automated} were among the first attempts to recognize bird species automatically by their sounds. They applied dynamic time warping and hidden Markov models for automatic song recognition of Zebra Finche and Indigo Punting. In~\cite{luque2016evaluation}, the authors focus on classifying two anuran species: Alytes obstetricans and Epidalea calamita using generic descriptors based on an MPEG-7 standard. Their evaluation demonstrate that MPEG-7 descriptors are suitable to be used in the recognition of different patterns, allowing a high scalability. In~\cite{hao2013monitoring}, the authors propose to classify animal sounds in the visual space, by treating the texture of animal sonograms as an acoustic fingerprint. Their method can obviate the complex feature selection process. They also show that by searching for the most representative acoustic fingerprint, they can significantly outperform other techniques in terms of speed and accuracy.

The WSN has been massively applied in sensing the environment and transferring collected samples to the server. However, it is challenging to realize in-network classification system because of the limited computational ability of wireless sensor node. Recently, several research works regarding in-network classification have been proposed. Sun et al.~\cite{sun2008dynamic} dynamically select the feature space in order to accelerate the classification process. A hybrid sensor networks is designed by Hu et al.~\cite{hu2009design} for in-network and energy-efficient classification in order to monitor amphibian population. Wei et al.~\cite{wei2013real} proposed a sparse representation classification method for acoustic classification on WSN. A dictionary reduction method was designed in order to improve the classification efficiency. The sparse representation classification method was also used by face recognition on resource-constrained smart phones to improve the classification performance~\cite{shen2014face,xu2016sensor}. 

Deep learning has achieved great success over the past several years for the excellent ability on high-level feature learning and  representative information discovering. Specifically, deep learning has been widely used in a number of areas, such as computer version \cite{wen2016discriminative}, activity recognition \cite{chen2018interpretable,luo2019brush}, sensory signal classification \cite{zhang2018multi,lan2019entrans,xu2019energy}, and brain computer interface \cite{zhang2018mindid}. Wen et al. \cite{wen2016discriminative} propose a new supervision signal, called center loss, for face recognition task. The proposed center loss function is demonstrated to enhance the discriminative power of the deeply learned features. Chen et al. \cite{chen2018interpretable} propose an interpretable parallel recurrent neural network with convolutional attentions to improve the activity recognition performance based on Inertial
Measurement Unit signals. Zhang et al. \cite{zhang2018multi} combine deep learning and reinforcement learning to deal with multi-modal sensory data (e.g., RFID, acceleration) and extract the latent information for better classification. Recently, deep learning involves in the brain signal mining in brain computer interface (BCI). Zhang et al. \cite{zhang2018mindid} propose an  attention-based Encoder-Decoder RNNs (Recurrent Neural Networks) structure in order to improve the robustness and adaptability of the brainwave based identification system. 

There are also several works that apply deep learning techniques in embedded devices. Lane et al.~\cite{lane2015can} propose low-power Deep Neural Network (DNN) model for mobile sensing. CPU and DSP in one mobile device are exploited for activity recognition. Lane et al.~\cite{lane2015can} also design a DNN model for audio sensing in mobile phone by using dataset from 168 places for the training purpose. A framework DeepX is further proposed for software accelerating on mobile devices~\cite{lane2016deepx}. 

In terms of animal voice classification, Zhang et al.~\cite{zhang2018automatic}, Oikarinen et al.~\cite{oikarinen2018deep} study animal voice classification using deep learning techniques. Our method is different from these two works. Our studies focus on voice classification in noisy environment while the voice data in~\cite{zhang2018automatic} are collected from controlled room without environmental noise. Instead of classifying different animals, \cite{oikarinen2018deep} analyses different call types of marmoset monkeys such as Trill, Twitter, Phee and Chatter. Moreover, we implement the proposed system on a testbed and evaluate its performance in real world environment. In another work~\cite{ntalampiras2018bird}, Stavros Ntalampiras used transfer learning to improve the accuracy of bird classification by exploiting music genres. Their result show that the transfer learning scheme can improve classification accuracy by $11.2\%$. Although the goal of this work and our study is to improve the recognition accuracy with deep learning technology, the methodologies are different. Our approach analyses the inherent features of audio signal and propose a multi-view CNN model to improve the accuracy. Instead of looking at the bird audio signals alone, Stavros Ntalampiras proposed to statistically analyse audio signals using their similarities with music genres. Their method, however, is only effective for a limited number of bird species because they need to perform feature transformation again when a new bird species comes in. In comparison, our approach is applicable for a large number of bird species. A number of studies also apply deep learning technologies in bird voice classification~\cite{potamitis2016deep,koops2014deep,goeau2016lifeclef}, however, they only use conventional deep learning approaches such as CNN and do not make any novel improvement. In this paper, we propose a multi-view CNN model and evaluation results show that the proposed model outperforms the conventional CNN.

\section{System Design}
\label{sec:system}

%In this section, we will introduce the architecture of the proposed system. 
\subsection{System Overview}
As shown in Figure~\ref{fig:systemoverview}, our proposed framework consists of two parts: WASN and server. In the WASN, the wireless nodes will detect and record animal voices and then perform local processing which include silence removal, segmentation and FFT. We process signal \emph{in-situ} before uploading because of the high sampling frequency of audio signal and energy inefficiency of wireless communication~\cite{barr2006energy,wei2013real}. The spectrum signal obtained from FFT can save half spaces since FFT is symmetric.
On the server side, the spectrum signal will be fed into a deep neural network to obtain the species of the animal. The classification results can be used by biologists to analyze the density, distribution and behavior of animals.

Wireless sensors are usually resource-poor relative to server, and not able to run computationally expensive algorithms such as deep learning models. Therefore, we assume all the wireless sensors can connect to a server via wireless communication technologies, such as ZigBee, Wi-Fi, and LoRa~\cite{xu2019measurement}. However, there may be network failure, server failure, power failure, or other disruption makes offload impossible. While such failures will hopefully be rare, they cannot be ignored in a cloud-based system. In this case, the node can transmit the data to the gateway or a nearby server which are usually resource-rich and capable of running deep learning models. Alternatively, the classification can be performed in the node to recognize only a few species, pre-defined by the user. When offloading becomes possible again, the system can revert to recognizing its full range of species. 

In the following parts, we will describe the design details of each component.

\subsection{Local Processing}
\textbf{Silence Removal.} The collected audio signal usually contains a large amount of silent signal when this is no animal present. Therefore, we apply a simple silence removal method on the raw signal to delete not-of-interest area. The procedure is explained in Algorithm~\ref{al:silence}. We first calculate the root mean square (RMS) of each window which contains 1s samples and then compare it with a pre-defined threshold learned from the environment. 
% If it is above the threshold, we keep it; otherwise, we delete it. 
The windows of samples whose RMS above the threshold will be kept. The threshold is determined by exhaustive search. To be specific, we increase the threshold from 0 to 0.5 with an increment of 0.01, then choose the one that can achieve the best performance (0.03 in this paper).
% The threshold is learned from the environment.

\begin{figure}[!ht]
    \centering
         \includegraphics[width=3in]{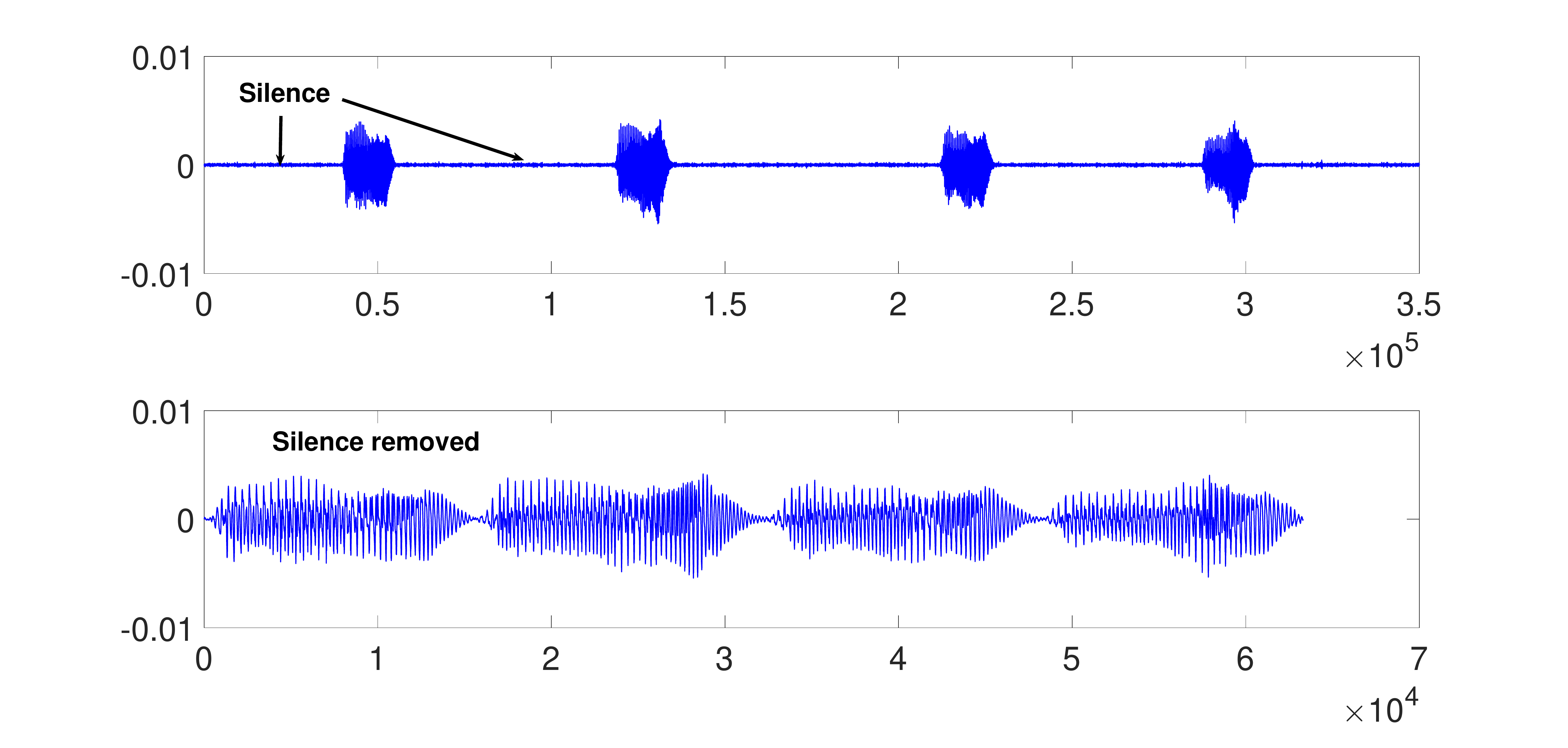}
    \caption{Silence removal.}
   \label{fig:silenceremoval}
\end{figure}

\begin{algorithm}[ht]
  \caption{Silence Removal}
  \label{al:silence}
  \begin{algorithmic}[1]
    \State{\textbf{Input}: Audio Segment $S_{i = 1:N} \in \mathbb{R} > 1$}, where $N$ is the total number of segments and $\rho$ is the threshold
    \For {$i = 1:N$}
      \If{RMS $(S_i) < \rho$}
        \State Remove $(S_i)$\;
      \EndIf
    \EndFor
  \end{algorithmic}
\end{algorithm}
Figure~\ref{fig:silenceremoval} shows an example of silence removal on an animal voice recording. We can see that it can effectively remove the silent periods and detect the present of animals.

\begin{figure*}[!ht]
    \centering
    \subfigure[Cultripes]{
         \includegraphics[width=1.6in,height=1.28in]{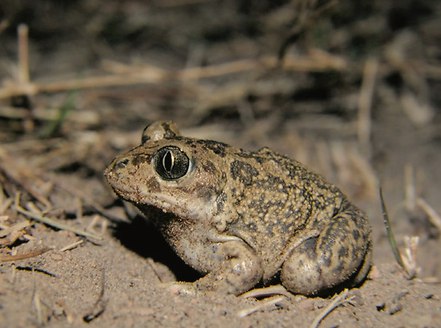}
        \label{fig:1}
    }
    \subfigure[Cultripes: time domain]{
         \includegraphics[width=1.6in,height=1.28in]{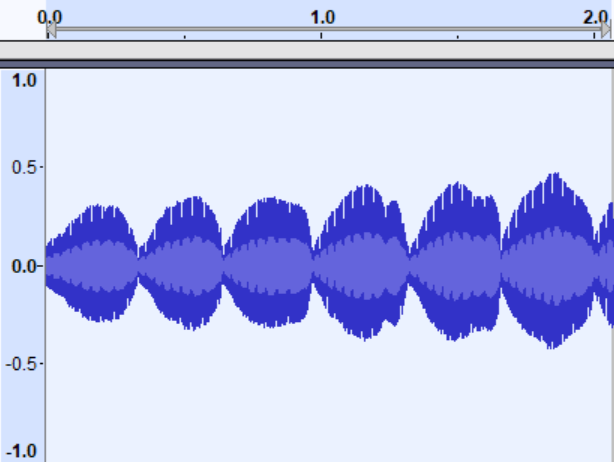}
        \label{fig:1}
    }
		 \subfigure[Cultripes: spectrum]{
         \includegraphics[width=1.6in]{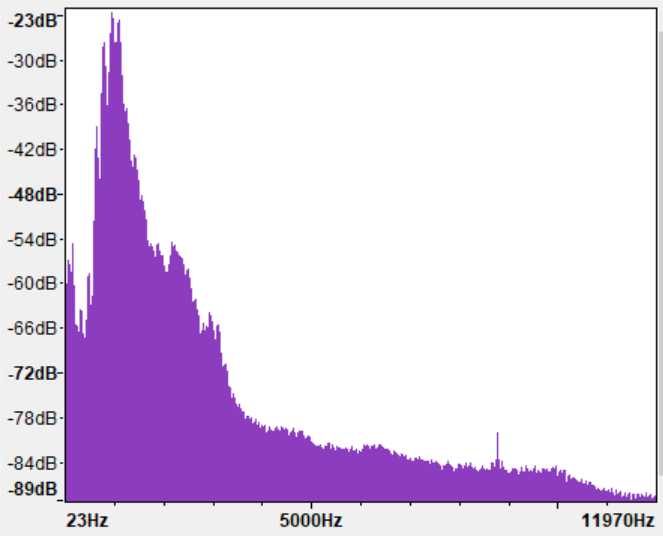}
      	 \label{fig:2}
   }
		\subfigure[Litoria Caerulea]{
         \includegraphics[width=1.6in,height=1.28in]{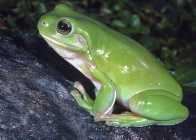}
        \label{fig:3}
    }
   \subfigure[Litoria Caerulea: time domain]{
         \includegraphics[width=1.6in,height=1.28in]{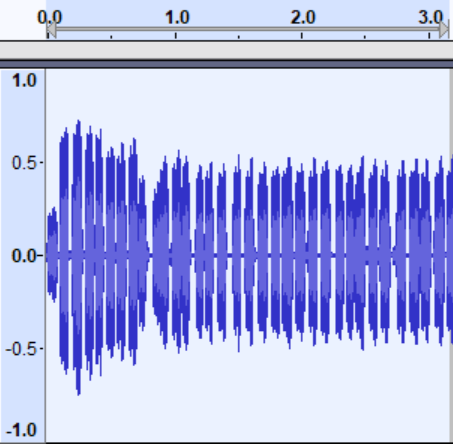}
        \label{fig:3}
    }
   \subfigure[Litoria Caerulea: spectrum]{
         \includegraphics[width=1.6in]{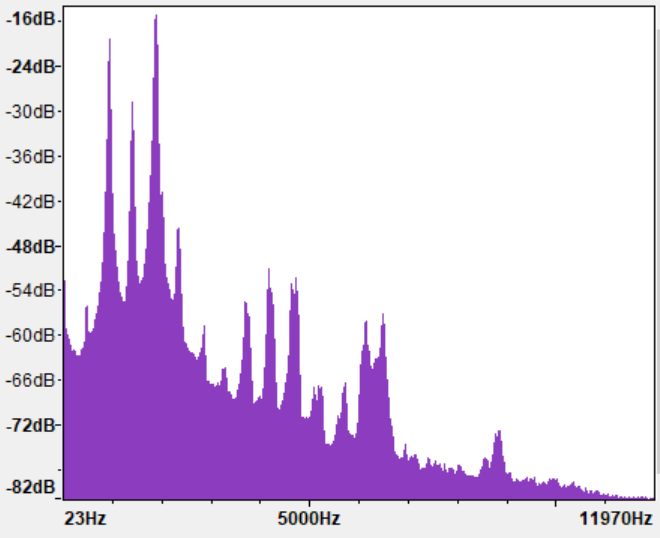}
      	 \label{fig:4}
   }
    \caption{Audio signal of two species of frog.}
   \label{fig:fft}
\end{figure*}

\textbf{Segmentation and FFT.} After silence removal, we obtain audio signals containing animal vocalization only. 
The audio signal is segmented into consecutive sliding windows with $50\%$ overlap. Hamming window is used in this paper to avoid spectral leakage. Each window contains $2^{14}$ samples which is chosen to balance the trade-off between classification accuracy and latency as discussed in Section~\ref{sec:evaluation}. The overlap in sliding window is used to capture changes or transitions around the window limits. Then we perform FFT on each segment to calculate spectrum energy (i.e. the magnitude of the FFT coefficients). As an example, Figure~\ref{fig:fft} shows the sound in time and frequency domain of two different frog species: Cultripes and Litoria Caerulea. It is conspicuous that they have different spectrum distributions. The graphs are plotted by audio signal analysis software Audacity.
\subsection{Multi-view Convolutional Neural Networks}
We propose a deep learning framework in order to automatically learn the latent and high-level features from the processed audio signals for better classification performance. Among deep learning algorithms, CNN is widely used to discover the latent spatial information in applications such as image recognition \cite{ciresan2011convolutional}, ubiquitous \cite{ning2018deepmag}, and object searching \cite{ren2017faster}, due to their salient features such as regularized structure, good spatial locality and translation invariance. CNN applies a convolution operation to the input, passing the result to the next layer. Specifically, CNN captures the distinctive dependencies among the patterns associated to different audio categories. However, one drawback of the standard CNN structure is that the filter length of the convolution operation is fixed. As a result, the convolutional filter can only discover the spatial features with the fixed filter range. For example, CNN may explore the short-term feature but fail to capture the middle- and long-term features. 
\begin{figure*}[!ht]
    \centering
         \includegraphics[width=5in]{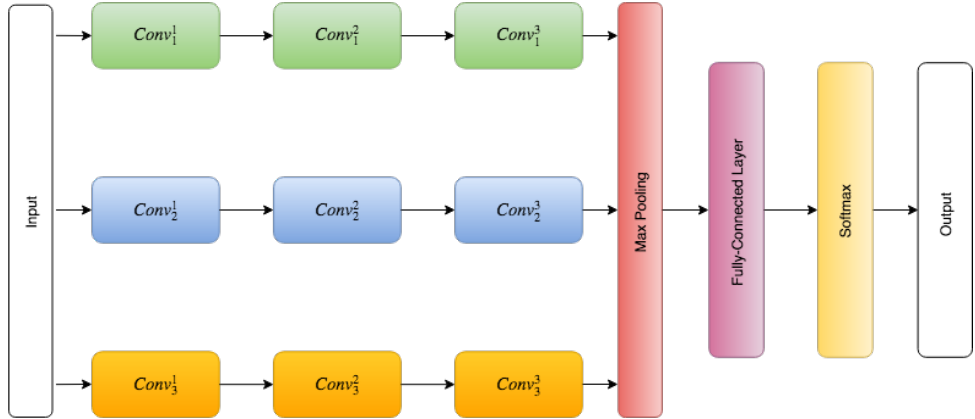}
    \caption{Multi-view CNN workflow.The audio signal is feed into three views for short-term, middle-term, and long-term spatial dependencies learning. $Conv_h^k$ denotes the $k$-th convolutional operation in the $h$-th view. The learned features from the multi-view structure are processed by the max pooling layer for dimension reduction, which are followed by the fully-connected layer, softmax layer, and at last predict the animal species as output.}
   \label{fig:CNN_workflow}
\end{figure*}

To address the mentioned challenge, we propose a multi-view CNN framework which applies three different filter length to extract the short-, middle-, and long-term features in parallel. As shown in Figure~\ref{fig:CNN_workflow}, the proposed framework regards the processed audio signals as input and feed into three views at the same time. Each view contains three convolutional layers. $Conv_h^k$ denotes the $k$-th convolutional operation in the $h$-th view. The convolutional layer contains a set of filters to convolve the audio data followed by the nonlinear transformation to extract the geographical features. 
The filter length keeps invariant in the same view while varies in different views. The extracted features from the multi-view pipe are stacked together and then through the max pooling operation for dimension reduction. Afterward, a fully-connected layer, a softmax layer and the output layer work as a classifier to predict the audio label. The proposed multi-view CNN has several key differences from the inception module~\cite{szegedy2015going} although the ideas are similar. First, ~\cite{szegedy2015going} has a $1 \times 1$ convolutional filter in the module in order to prevent the information corruption brought by inter-channel convolutions. The proposed multi-view CNN does not has this component. This is because in our case the input data are naturally formed as a vector which represents the spectral information of the acoustic signals.  Moreover, ~\cite{szegedy2015going} adds an alternative parallel pooling path in the middle layer to acquire additional beneficial effect. However, we believe this may cause information loss and only perform the pooling operation after the concentration of the results of various views.
 
 \begin{table*}[h]
\caption{Species used in the experiments.}
\label{tab:species}
\centering
\small
\resizebox{6.5in}{!}{
\begin{tabular}{|c|c|c|c|}
\hline
\multicolumn{2}{|c|}{Frog dataset}                   & \multicolumn{2}{c|}{Cricket dataset (1 belongs to Gryllidae, 2 belongs to Tettigoniidae)}                 \\ \hline
Cyclorana Cryptotis          & Cyclorana Cultripes   & Acheta$^{1}$          & Aglaothorax$^{2}$   \\ \hline
Limnodynastes Convexiusculus & Litoria Caerulea      &Allonemobius$^{1}$ & Amblycorypha$^{2}$      \\ \hline
Litoria Inermis              & Litoria Nasuta        & Anaxipha$^{1}$              & Anaulacomera$^{2}$        \\ \hline
Litoria Pallida              & Litoria Rubella       & Anurogryllus$^{1}$              & Arethaea$^{2}$        \\ \hline
Litoria Tornieri             & Notaden Melanoscaphus & Cyrtoxipha$^{1}$            & Atlanticus$^{2}$  \\ \hline
Ranidella Bilingua           & Ranidella Deserticola & Eunemobius$^{1}$           & Belocephalus$^{2}$  \\ \hline
Uperoleia Lithomoda          & Bufo Marinus          & Gryllus$^{1}$          & Borinquenula$^{2}$          \\ \hline
                             &                       &Hapithus$^{1}$                           & Bucrates$^{2}$                     \\ \hline
                             &                       &Capnobotes$^{2}$                           & Caribophyllum$^{2}$                     \\ \hline
                             &                       & Ceraia$^{2}$                            & Conocephalus$^{2}$                     \\ \hline
\end{tabular}
}
\end{table*}
Suppose the input audio data $\bm{E}$ has shape $[M, L]$ with depth as $1$. The chosen three convolutional filters with size in short-, middle-, and long-term views are $[M, 10]$, $[M, 15]$, $[M, 20]$, respectively. The stride sizes keep $[1, 1]$ for all the convolutional layers. The stride denotes the x-movements and y-movements distance of the filters. Since the audio signals are arranged as 1-dimension data, we set $M = 1$.  Same shape zero padding is used, which keeps the sample shape constant during the 
the convolution calculation. In the convolutional operation, the feature maps from the input layer are convolved with the learnable filters and fed to the activation function to generate the output feature map. For a specific convolutional area (also called perceptive area) $\bm{x}$ which has the same shape as the filter, the convolutional operation can be described as
$$\bm{x}' = tanh(\sum_{i}\sum_{j}\bm{f}_{ij}*\bm{x}_{ij})$$
where $\bm{x}'$ denotes the filtered results while $\bm{f}_{ij}$ denotes the $i$-th row and the $j$-th column element in the trainable filter. We adopt the widely used \textit{tanh} activation function for nonlinearity.
The depth of input sample transfers to $D$ through the convolutional layer and the sample shape is changed to $[M, L, D]$. In particular, the corresponding depth $D_h= 2, 4, 8$ for three convolutional layers. The features learned from the filters are concatenated and flattened to $[1, M*L*\sum_{h=1}^{3}D_h]$. The max pooling has $[1, 3]$ as both pooling length and strides. Therefore, the features with shape $[1, M*L*\sum_{h=1}^{3}D_h/3]$ after the pooling operation, which are forwarded to the fully-connected layer. The operation between the fully-connected layer and the output layer can be represented by 
$$\bm{y} = softmax(\bar{\bm{w}}\bm{E^{FC}} + \bar{\bm{b}})$$
where $FC$ denotes the fully-connected layer while the $\bar{\bm{w}}$ and $\bar{\bm{b}}$ denote the corresponding weights matrix and biases.
The softmax function is used for activation. For each sample, the corresponding label information is presented by one-hot label $\bm{y} \in \mathbb{R}^H$ where $H$ denotes the category number of acoustic signals. The error between the predicted results and the ground truth is evaluated by cross-entropy
$$loss = - \sum_{h=1}^{H}\bm{y}_h log(p_h)$$
where $p_h$ denotes the predicted probability of observation of an object belonging to category $h$.
The calculated error is optimized by the AdamOptimizer algorithm~\cite{kingma2014adam}. To minimize the possibility of overfitting, we adopt the dropout strategy and set the drop rate to 80\%. 

\begin{figure*}[!ht]
    \centering
    \subfigure[Impact of window size.]{
         \includegraphics[width=2in]{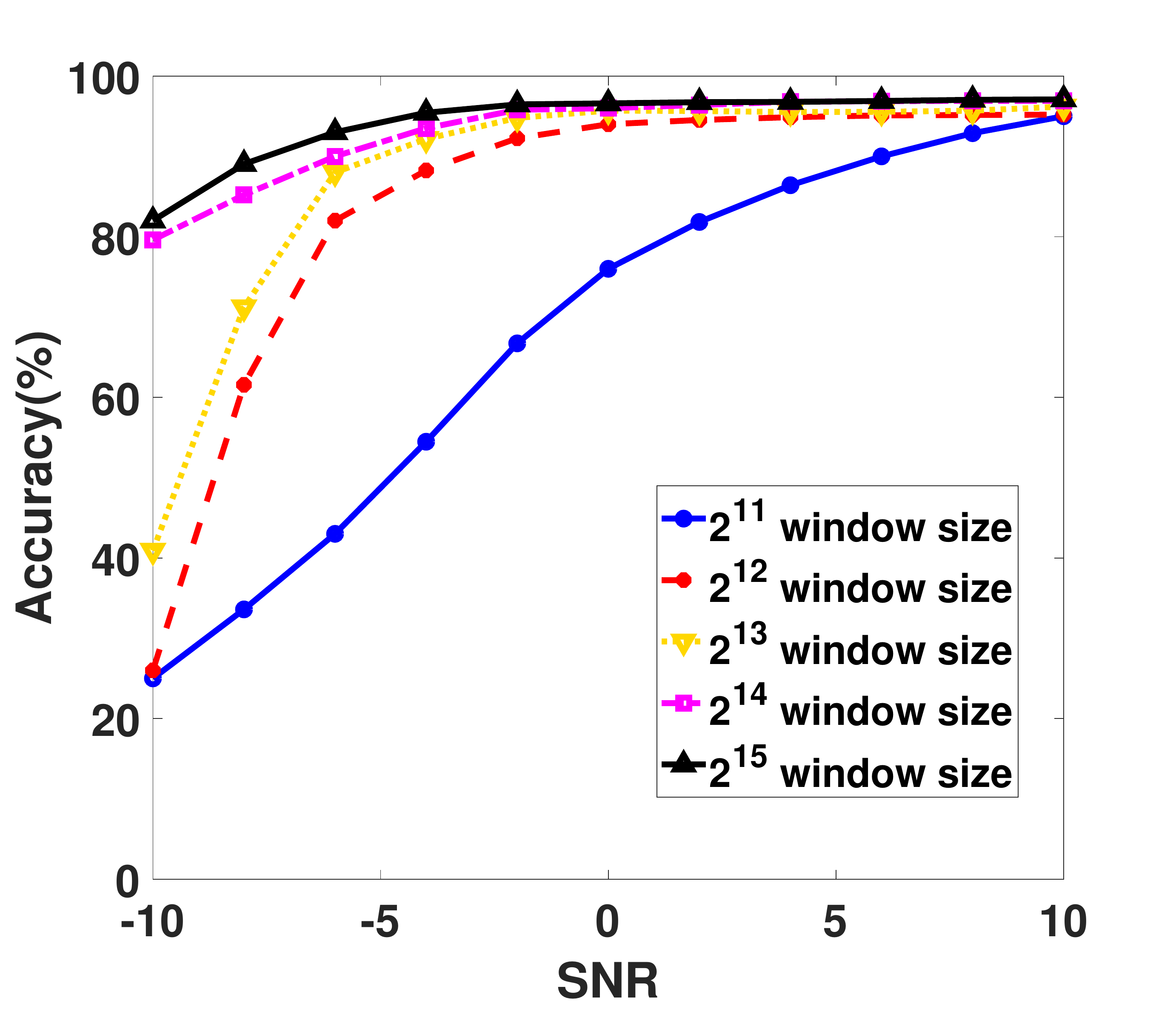}
        \label{fig:impactofwindow_Frog}
    }
    \subfigure[Impact of iterations.]{
         \includegraphics[width=2in]{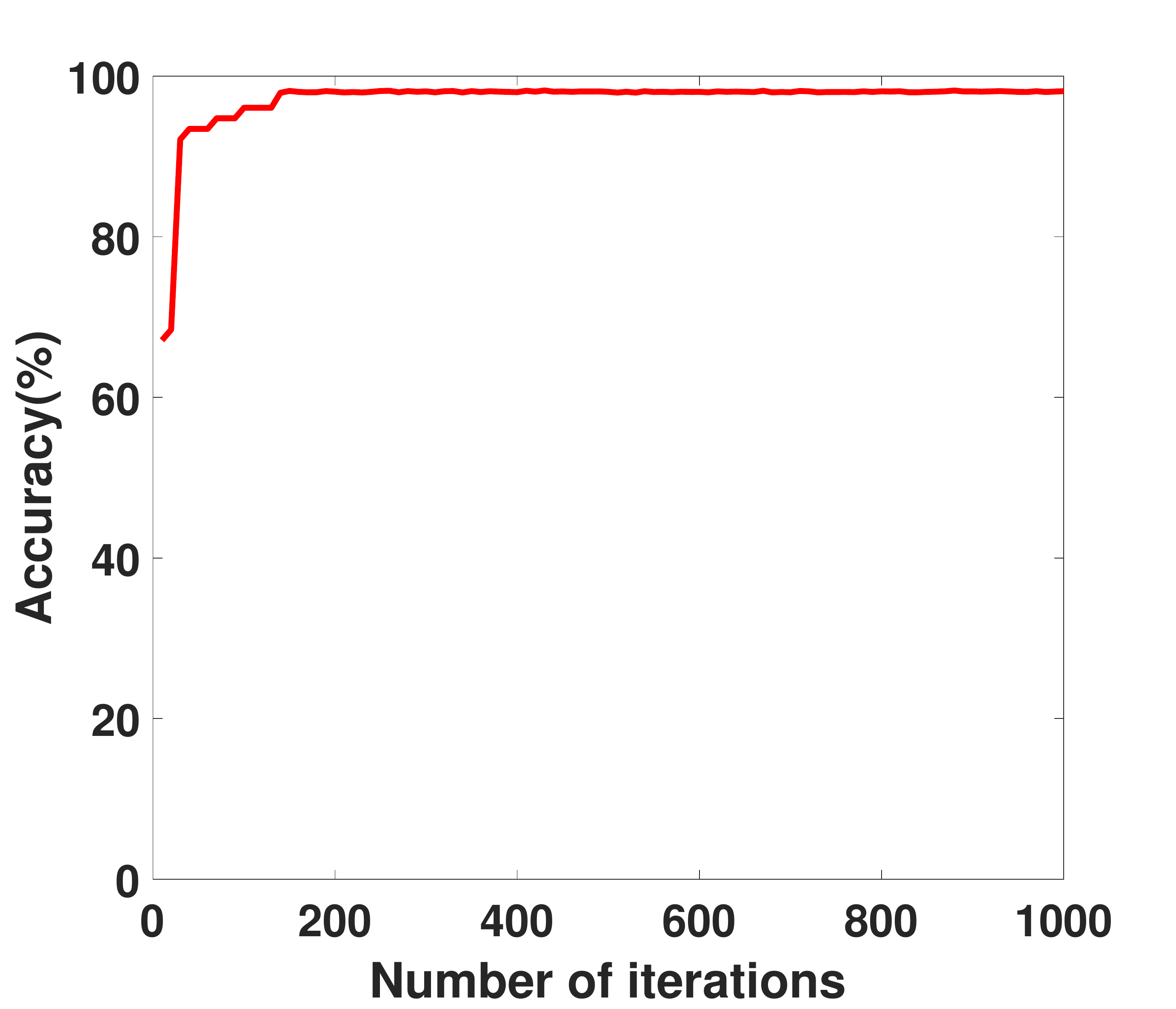}
        \label{fig:impactofiteration_Frog}
    }
    \subfigure[Impact of dropout rate.]{
         \includegraphics[width=2in,height=1.71in]{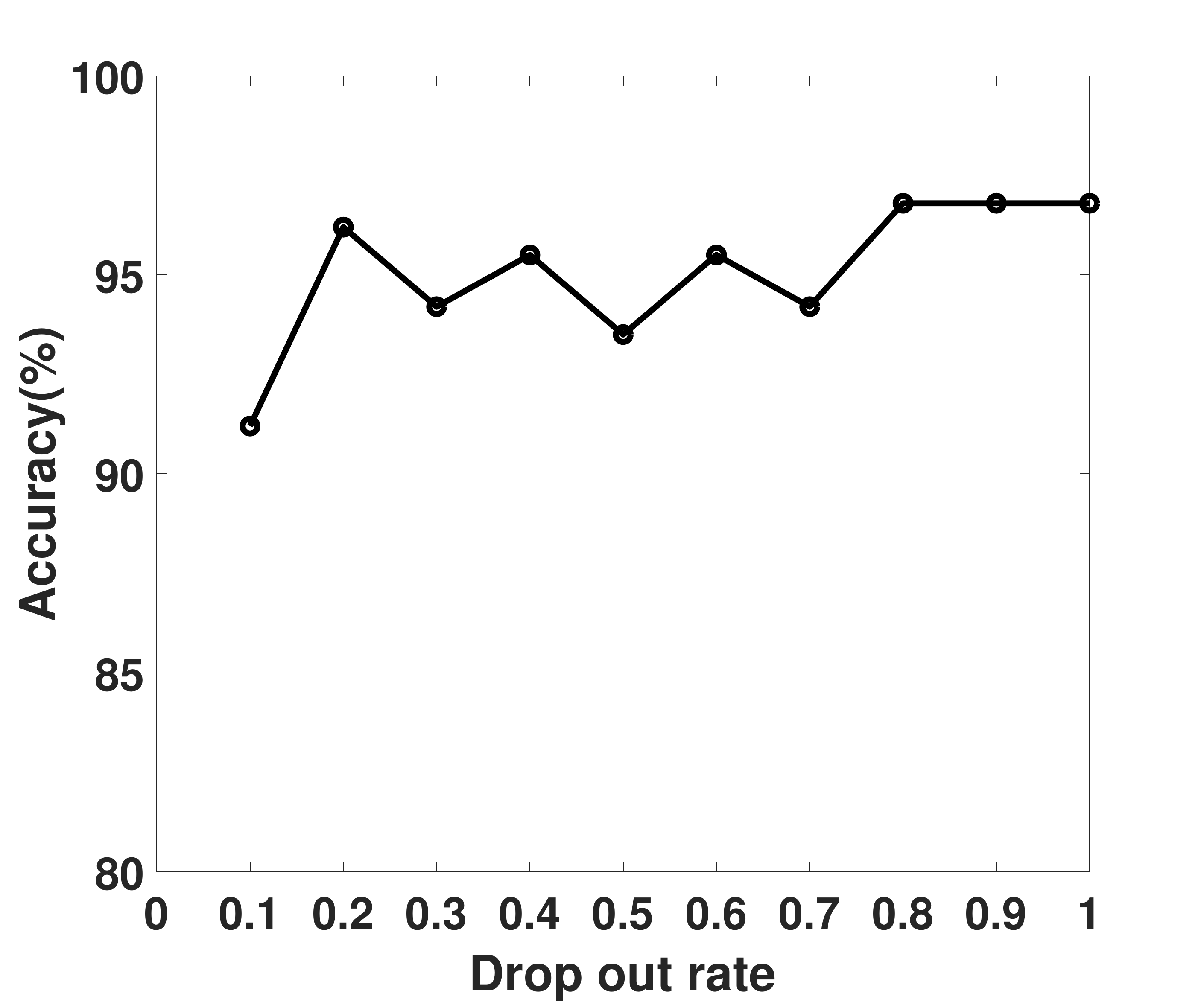}
        \label{fig:impactofdropout_Frog}
    }
    \subfigure[Impact of learning rate.]{
         \includegraphics[width=2in,height=1.71in]{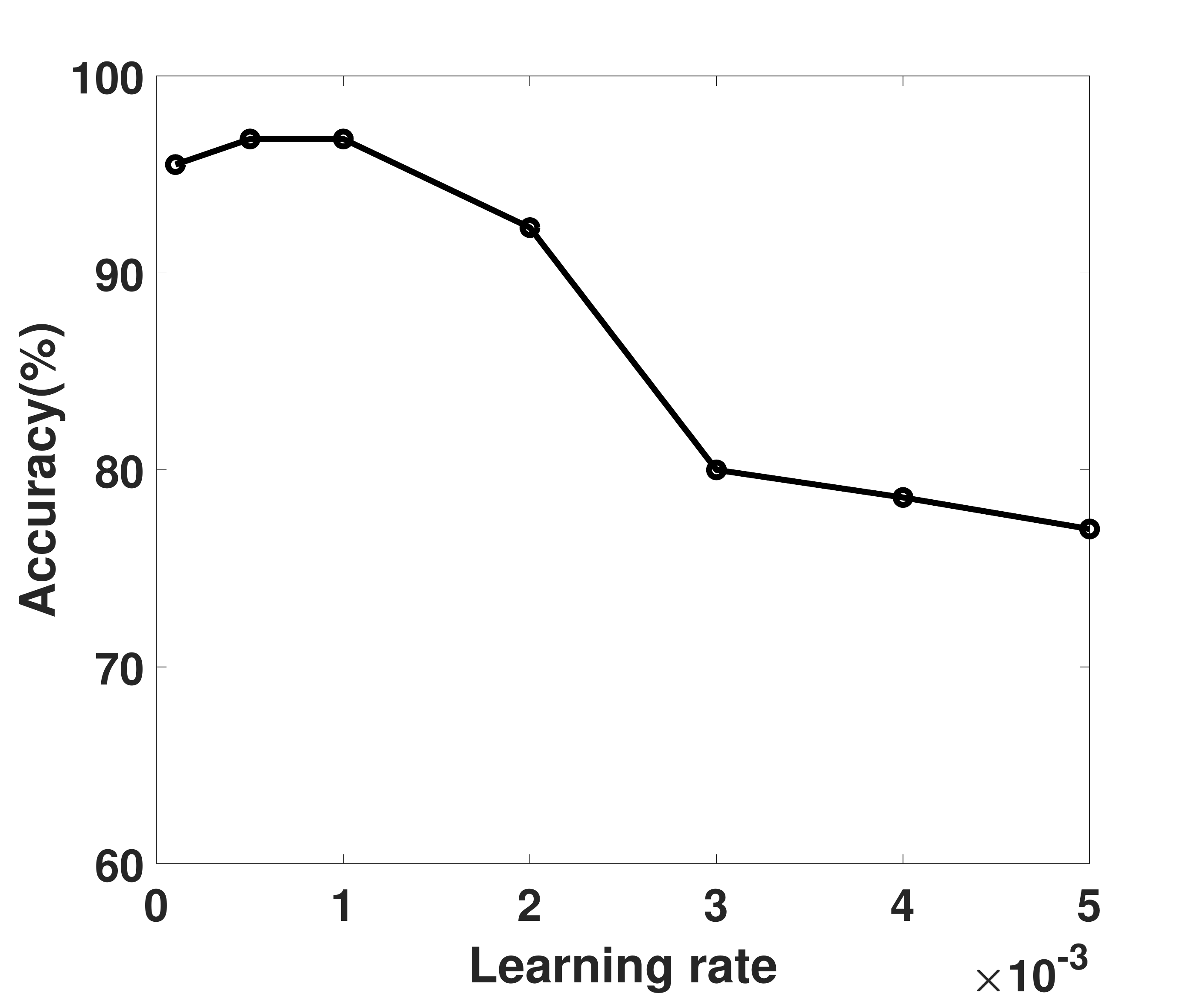}
        \label{fig:impactoflearningrate_Frog}
    }
    \subfigure[Impact of training dataset.]{
         \includegraphics[width=2in,height=1.71in]{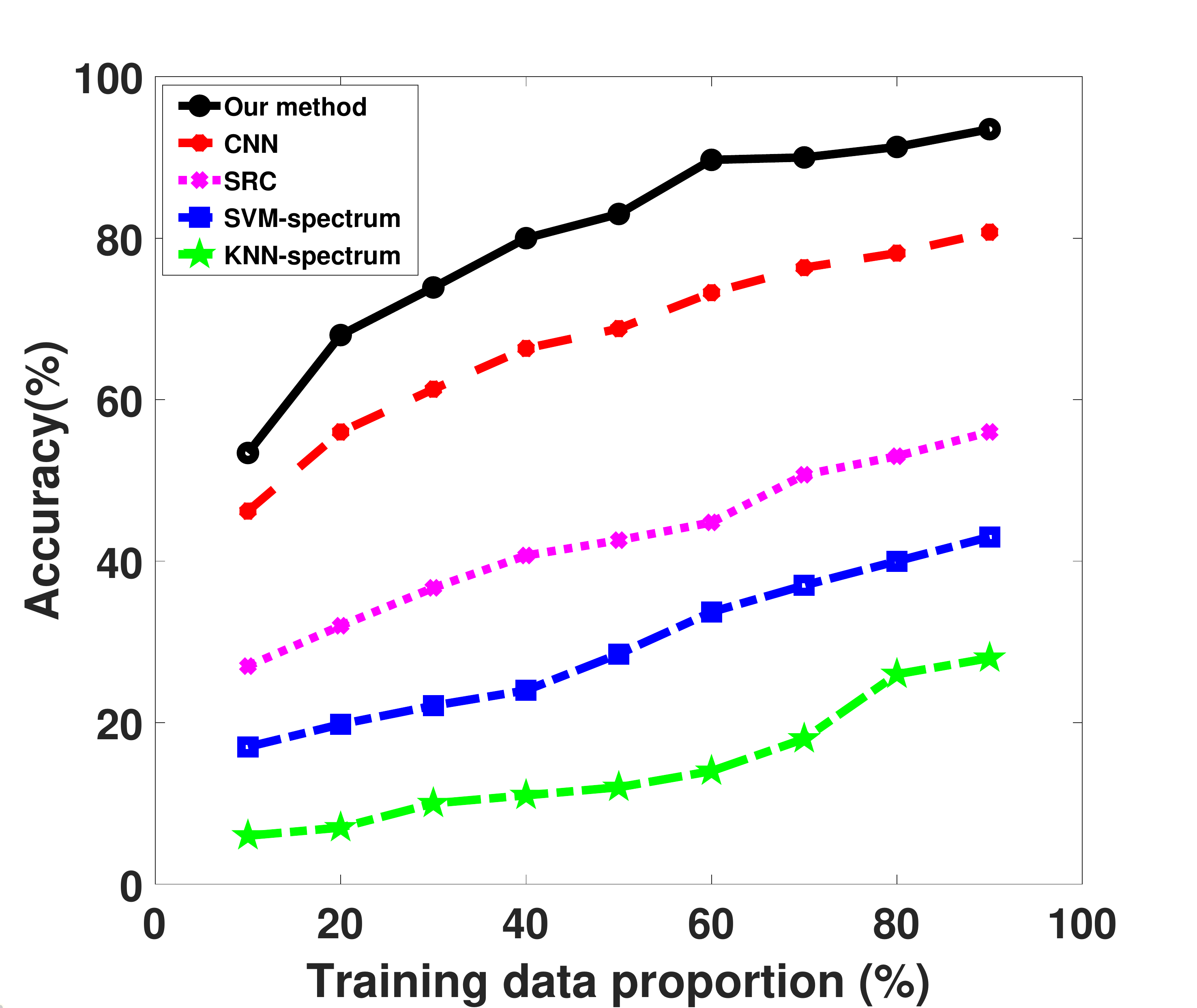}
        \label{fig:impactoftrainingsize_Frog}
    }
    \subfigure[Comparison with other methods.]{
         \includegraphics[width=2in,height=1.71in]{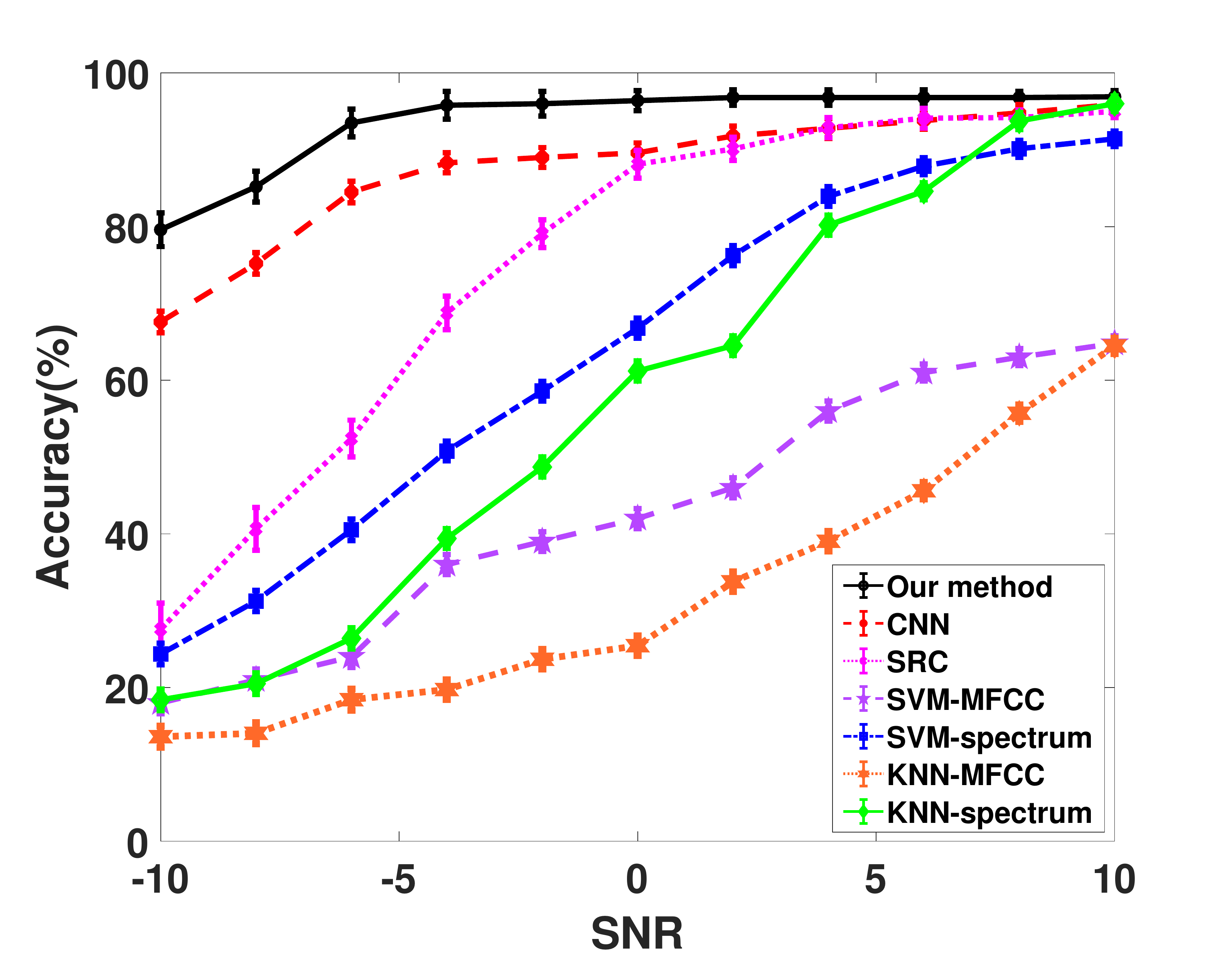}
        \label{fig:comparisonwithothers_Frog}
    }
    \caption{Evaluation results of frog dataset.}
   \label{fig:frogdataset}
\end{figure*}
\section{Evaluation}
\label{sec:evaluation}
\subsection{Goals, Metrics, and Methodology}
\label{subsec:goals}
In this section, we evaluate the performance of the proposed system based on two real datasets. The goals of the evaluation are twofold: 1) evaluate the performance of the proposed system under different settings; 2) compare the proposed system with previous animal vocalization system. 
\begin{figure}[!ht]
    \centering
         \includegraphics[width=3.2in]{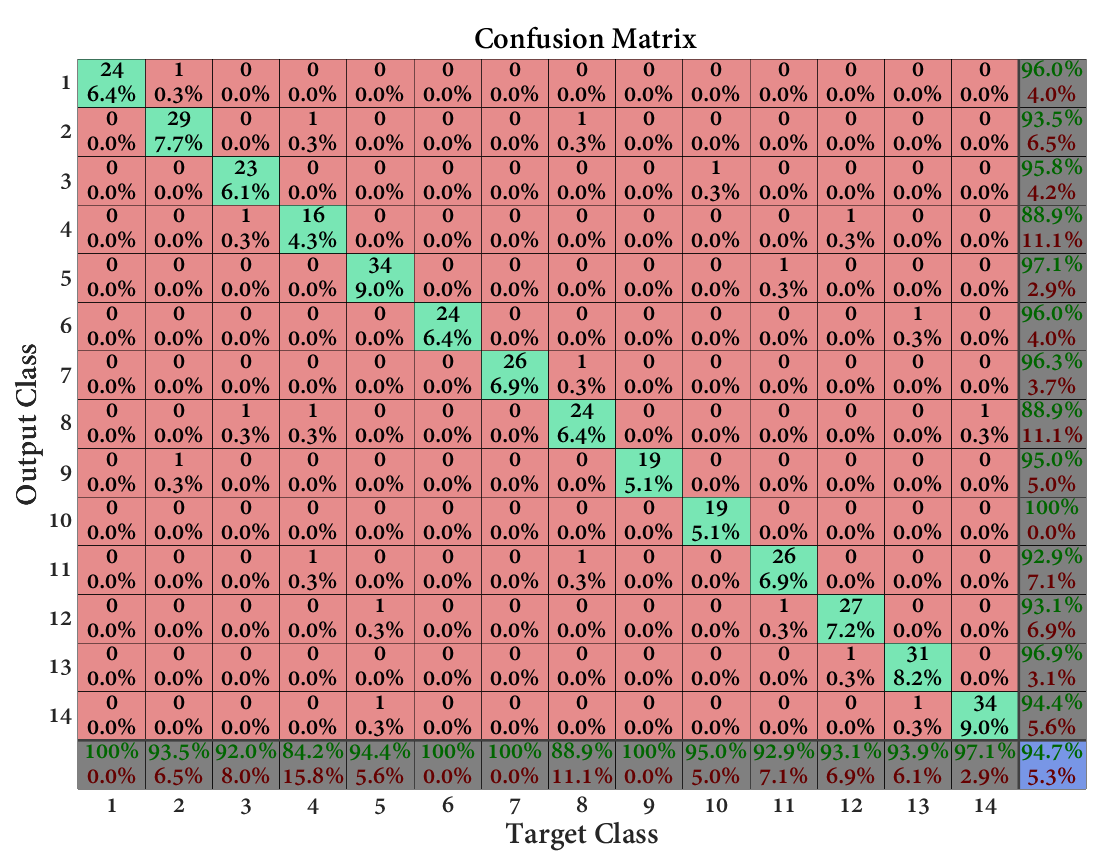}
  \caption{Confusion matrix of frog dataset.}
  \label{fig:confusionmatrix_frog}
\end{figure}
We use two datasets collected from real-world for evaluation. The first dataset contains audio signals recorded from fourteen different species of frogs. The sampling frequency for this dataset is 24Khz. More details about this dataset can be found in~\cite{wei2013real}. The second dataset~\footnote{http://alumni.cs.ucr.edu/~yhao/animalsoundfingerprint.html} contains audio signals recorded from different species of crickets. The data consists of twenty species of crickets, eight of which are Gryllidae and twelve of which are Tettigoniidae. The sampling frequency is also 24Khz. More details about this dataset can be found in~\cite{hao2013monitoring}. For completeness, Table.~\ref{tab:species} lists all the species we used in the experiments.
\begin{table*}[h]
\caption{Performance of different methods on frog dataset (SNR=-6dB).}
\label{tab:comparison_frog}
\centering
\small
\begin{tabular}{cccccccc}
\toprule
          & Our method        &CNN        & SRC & SVM-MFCC & SVM-Spectrum & KNN-MFCC & KNN-Spectrum \\ \hline
Accuracy  & \textbf{94.7\%}   &82.7\%      & 53.4\% & 24.4\%      & 40.5\%          & 20.1\%      & 26.4\%          \\ \hline
Precision &  \textbf{93.1\%}  &81.6\%       & 54.2\% & 25.9\%      & 43.5\%          & 19.9\%      & 25.1\%          \\ \hline
Recall    &  \textbf{94.3\%}  &82.4\%       & 53.7\% & 24.7\%      & 41.2\%          & 21.5\%      & 27.1\%          \\ \hline
F1-score  &  \textbf{92.9\%}  &81.2       & 52.1\% & 25.1\%      & 39.6\%          & 20.7\%      & 25.7\%          \\ \bottomrule
\end{tabular}
\end{table*}
In this paper, we use SVM and KNN to benchmark ASN classification because they have been widely used in WASN classification systems~\cite{fagerlund2007bird,huang2009frog,acevedo2009automated}. We evaluate the performance of SVM and KNN by using frequency domain and Mel-frequency cepstral coefficients (MFCCs), respectively. The parameters in SVM and KNN are well tuned to give highest accuracy. In addition, we compare the accuracy of our system with a recent work which is based on SRC~\cite{wei2013real} and conventional CNN. In total, we compare our method with six classifiers: CNN, SRC, SVM-MFCC, SVM-spectrum, KNN-MFCC and KNN-spectrum. For each classifier, we perform 10-fold cross-validation on the collected dataset. In the original dataset, the data only contain little environment noise. Therefore, to demonstrate the robustness of the proposed framework, we add different scales of environmental noise to create different SNRs. This is used to simulate the real environment because the recorded animal voices are usually deteriorated by environmental noise in real WASN. In the evaluation, it is done by adding different scales of random Gaussian noise to the original audio data.
\begin{figure*}[!ht]
    \centering
    \subfigure[Impact of window size.]{
         \includegraphics[width=2in]{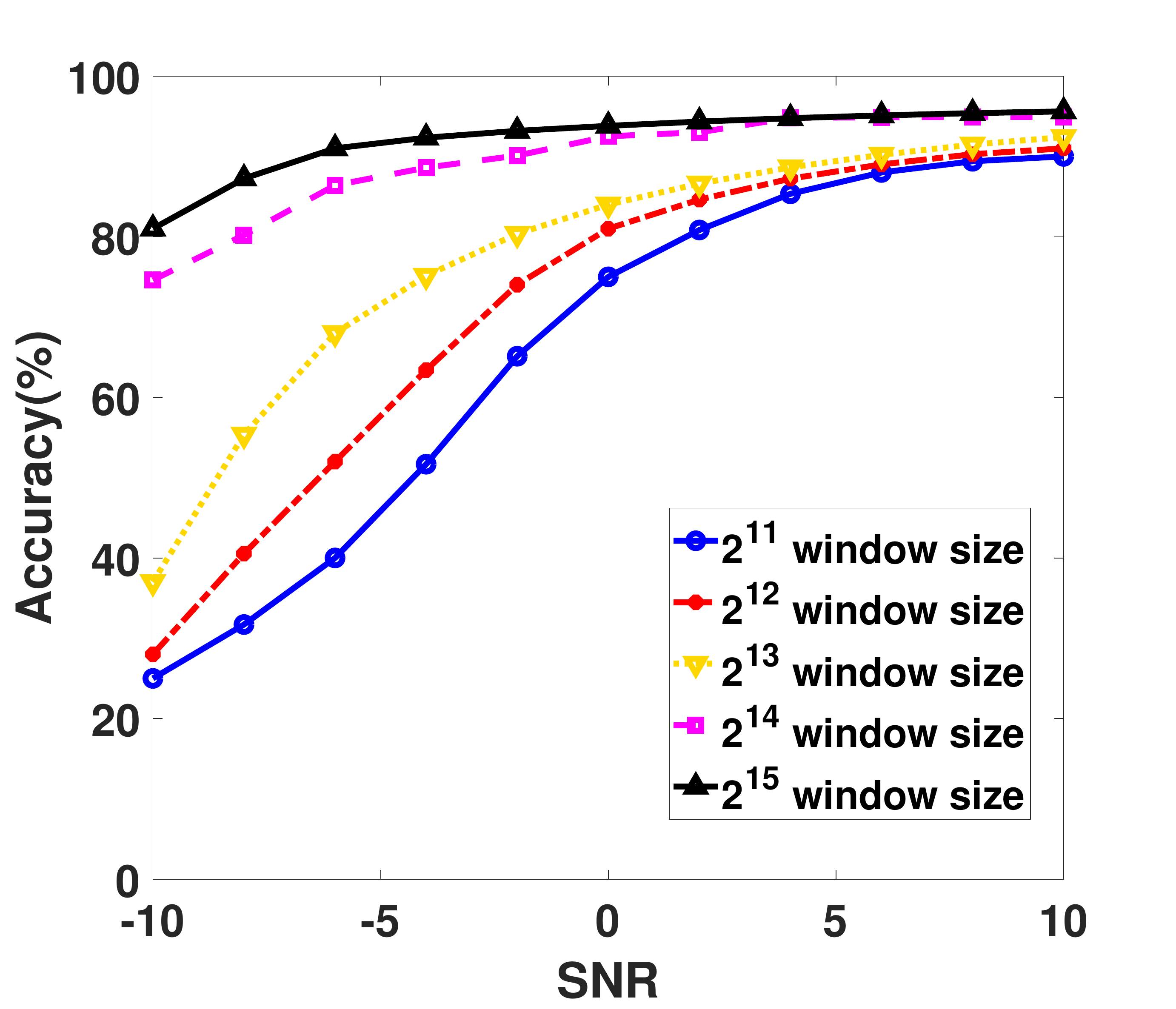}
        \label{fig:impactofwindow_cricket}
    }
    \subfigure[Impact of iterations.]{
         \includegraphics[width=2in]{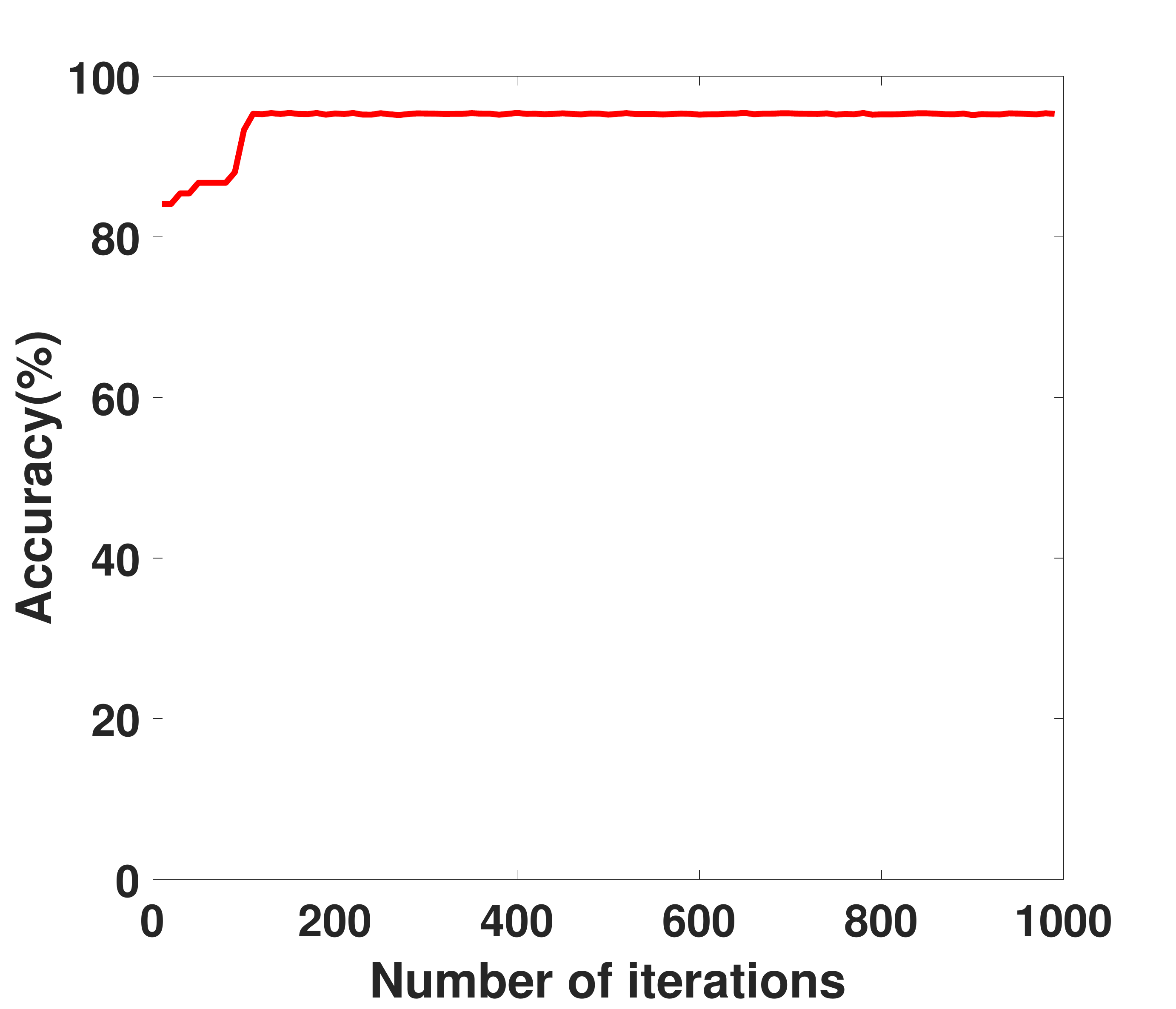}
        \label{fig:impactofiteration_cricket}
    }
   \subfigure[Impact of dropout rate.]{
         \includegraphics[width=2in,height=1.71in]{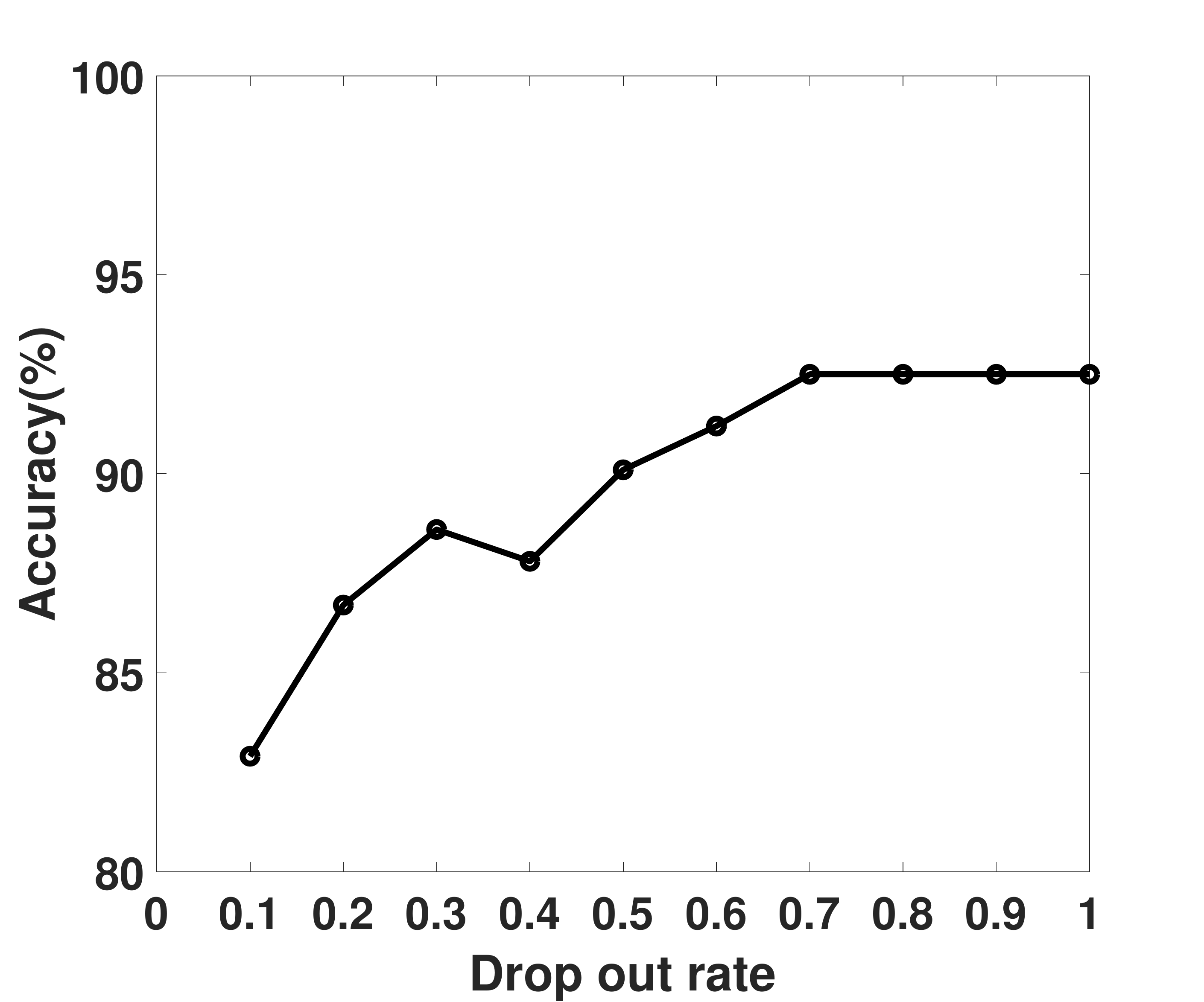}
        \label{fig:impactofdropout_cricket}
    }
    \subfigure[Impact of learning rate.]{
         \includegraphics[width=2in,height=1.71in]{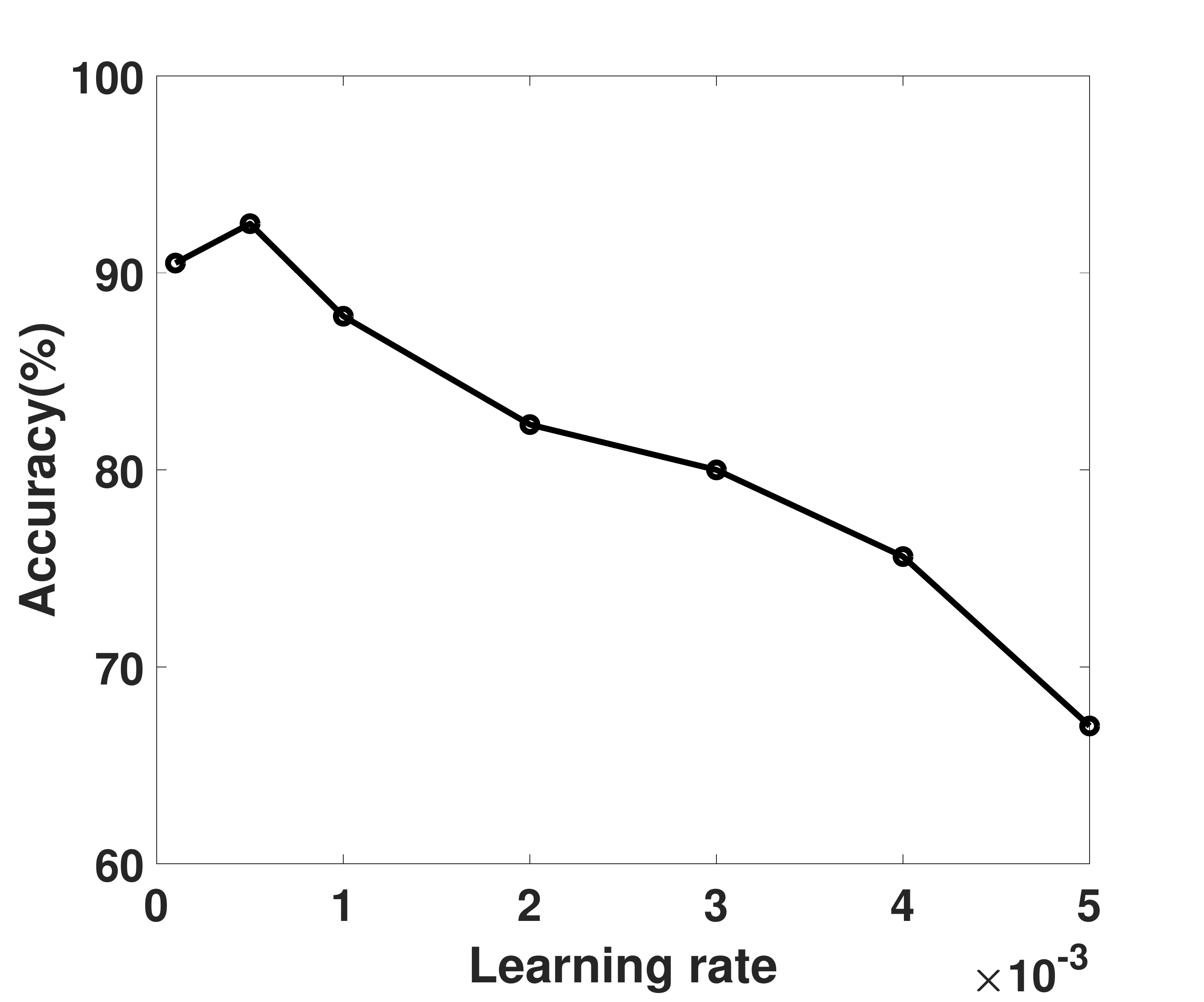}
        \label{fig:impactoflearningrate_cricket}
    }
    \subfigure[Impact of training dataset.]{
         \includegraphics[width=2in,height=1.71in]{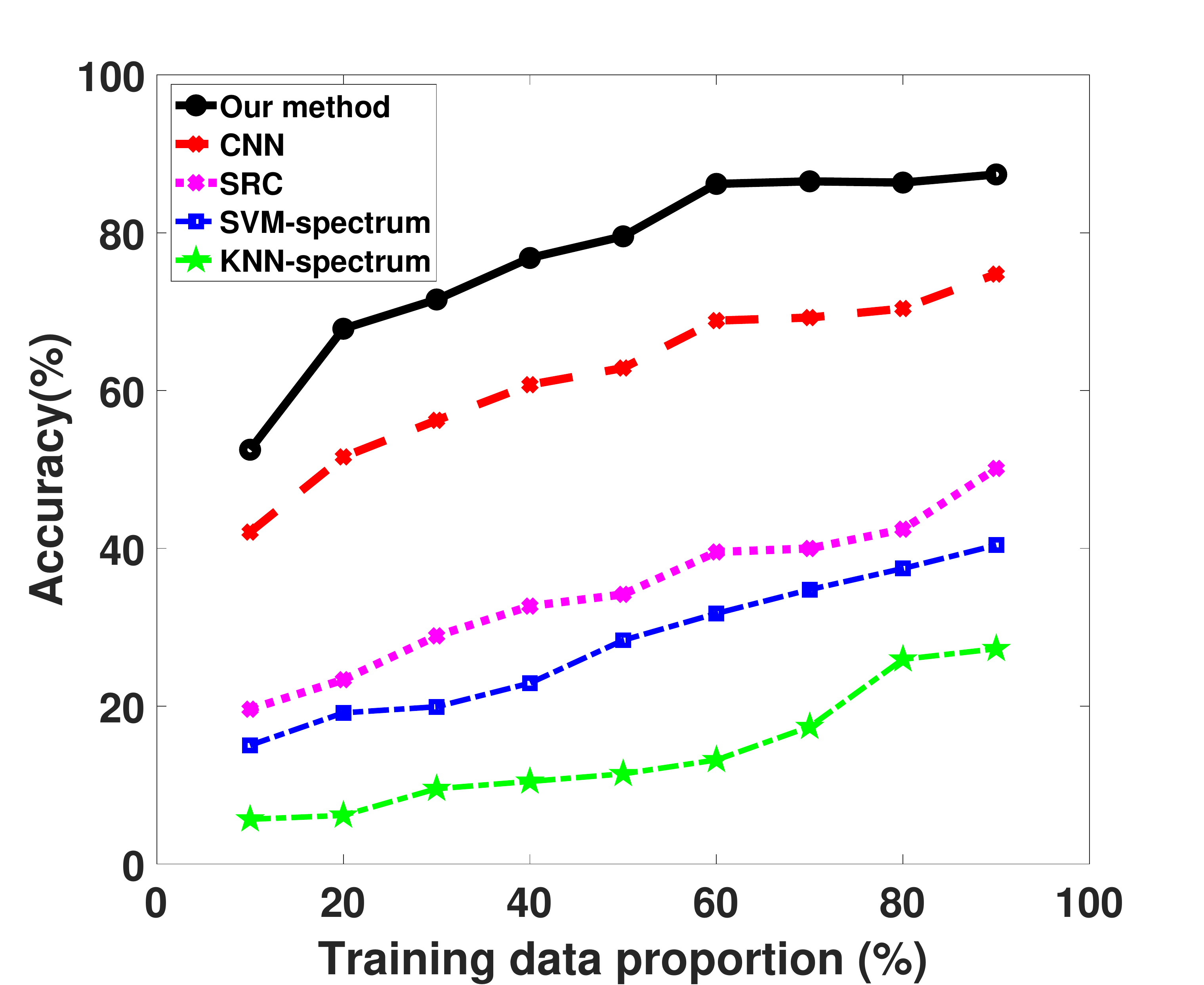}
        \label{fig:impactoftrainingsize_cricket}
    }
    \caption{Evaluation results of cricket dataset.}
   \label{fig:cricketdataset}
\end{figure*}
In this paper, we focus on the following four metrics: \textit{accuracy}, \textit{precision}, \textit{recall} and \textit{F1-score}. We plot the results of the average values and stand deviation obtained from 10 folds cross-validation.

\subsection{Performance of Frog Dataset}
\subsubsection{Impact of parameters}
We first evaluate the impact of important parameters in our system. On the node's side, the important parameters include window size of segment. On the server's side, the important parameters include the number of iterations in training, the dropout rate and learning rate in CNN, the size of training dataset. Dropout is a technique where randomly selected neurons are ignored during training. For example, the dropout rate of $80\%$ means that we randomly select $20\%$ of the neurons and drop them (force the values of them as 0). The dropout strategy is widely used to enhance the generalization of a machine learning model and prevent overfitting. The learning rate is a hyper-parameter that controls how much we are adjusting the weights of the neuron network with respect to the loss gradient.

To evaluate the impact of window size, we vary the window size from $2^{11}$ to $2^{15}$ samples and calculate the accuracy of our scheme. From the results in Figure~\ref{fig:impactofwindow_Frog}, we can see that there is a performance gain when we increase the window size and the improvement reduces after $2^{14}$ samples. Although we can achieve higher accuracy with more samples, the resource consumption of FFT operation which runs on the wireless sensor node also increases. Therefore, we choose to use $2^{14}$ window size to balance the trade-off between accuracy and resource consumption.

Figure~\ref{fig:impactofiteration_Frog} shows the accuracy along with different training iterations. We can see that the proposed method converges to its highest accuracy in less than 200 iterations. The results show that the proposed framework can finish training quickly. Figure~\ref{fig:impactofdropout_Frog} plots the accuracy of various dropout rates. We can observe that the accuracy fluctuates first and then becomes stable after the dropout rate is greater than 0.8. Therefore, we set the default dropout rate to be 0.8. Moreover, we can infer from Figure~\ref{fig:impactofdropout_Frog} that our model is not very sensitive to the dropout rate. This is because the Frog dataset matches well with the proposed multi-view CNN, as a result, the convergence suffers less from overfitting which can be demonstrated by the good convergence property as shown in Figure~\ref{fig:impactofiteration_Frog}. Figure~\ref{fig:impactoflearningrate_Frog} shows the accuracy under different learning rates. We can see that it achieves the highest accuracy when the learning rate is $0.5\times 10^{-3}$ and $1 \times 10^{-3}$. Correspondingly, we choose 0.001 to reduce the training time because the smaller the learning rate is, the slower the training process is. From Figure~\ref{fig:impactoflearningrate_Frog}, we can observe that the performance varies dramatically with the increasing of learning rate. One possible reason for this is that the gradient surface of our loss function is not smooth and very sensitive to the learning rate. The optimiser is easy to step over the local optima while the learning rate is larger than a threshold.

Next, we evaluate the accuracy of the proposed system under different sizes of training dataset. In this experiment, we use different proportions of the whole dataset for training, and use the left dataset for testing. The proportion increases from $10\%$ to $90\%$ with an increment of $10\%$. For example, the proportion of $10\%$ means we use $10\%$ of the dataset for training, and use the left dataset for testing. For comparison purpose, we also calculate the accuracy of CNN, SRC, SVM and KNN. From the results in Figure~\ref{fig:impactoftrainingsize_Frog}, we can see that our method continuously achieves the highest accuracy, and the accuracy becomes relatively stable after $60\%$ of the dataset is used for training. We also notice that the improvement of our method from $10\%$ to $90\%$ is remarkable. More specifically, when the proportion of the training dataset increases from $10\%$ to $90\%$, the accuracy improvement of our method is $40.1\%$ while the improvement of CNN, SRC, SVM and KNN are $34.7\%$, $29.3\%$, $26.7\%$ and $22.4\%$, respectively. In this experiment, we do not test SVM-MFCC and KNN-MFCC because their accuracy is poor as will be shown later.

\begin{table*}[!h]
\caption{Performance of different methods on cricket dataset (SNR=-6dB).}
\label{tab:comparison_cricket}
\centering
\small
\begin{tabular}{cccccccc}
\toprule
          & Our method        &CNN    & SRC & SVM-MFCC & SVM-Spectrum & KNN-MFCC & KNN-Spectrum \\ \hline
Accuracy  &  \textbf{86.4\%}  &76.6\%      & 42.4\% & 22.1\%      & 36.8\%          & 19.4\%      & 28.5\%          \\ \hline
Precision & \textbf{86.9\%}    &76.2\%     & 42.5\% & 23.6\%      & 36.3\%          & 18.2\%      & 28.6\%          \\ \hline
Recall    & \textbf{85.1\%}   &75.3\%     & 41.2\% & 22.7\%      & 37.3\%          & 19.9\%      & 29.8\%          \\ \hline
F1-score  & \textbf{86.1\%}   &74.6\%     & 41.7\% & 21.8\%      & 38.1\%          & 19.5\%      & 29.5\%          \\ \bottomrule
\end{tabular}
\end{table*}
\subsubsection{Comparison With Other Methods}
We now compare the performance of proposed scheme with previous approaches. As mentioned above, we compare the accuracy of the proposed system with conventional CNN, SRC, SVM-MFCC, SVM-spectrum, KNN-MFCC and KNN-spectrum. The MFCC of each window is calculated by transforming the power spectrum of each window into the logarithmic mel-frequency spectrum. We calculate the accuracy of different methods under different SNRs by adding different scales of environmental noise.  

As we can see from Figure~\ref{fig:comparisonwithothers_Frog}, SVM-MFCC and KNN-MFCC performs the worst which suggests that different frog species are not distinguishable in MFCC feature space. The results also explains why MFCC-based methods usually requires other carefully selected features~\cite{vaca2010using}. We find that when the animal voice is overwhelmed by environmental noise (low SNR), the accuracy of our system is significantly higher than the other methods. For example, when $SNR=-6dB$, the accuracy of our method is $12\%$ higher than CNN, $41\%$ higher than SRC, $70\%$ higher than SVM-MFCC, $53.9\%$ higher than SVM-spectrum, $74.3\%$ higher than KNN-MFCC, and $68\%$ higher than KNN-spectrum. The robustness to noise makes the proposed system suitable for real deployment in noisy environments. Moreover, the results also indicate that our system needs less sensors to cover a certain area because our system can classify low SNR signals which are usually collected from longer distance.

To take a closer look at the result, we summarize the results of different methods in Table~\ref{tab:comparison_frog} and plot confusion matrix in Figure~\ref{fig:confusionmatrix_frog} when SNR is -6dB. We can see that each class can achieve high accuracy and the overall average accuracy is $94.7\%$. 
\begin{figure}[!ht]
    \centering
    \subfigure[Two-class classification.]{
         \includegraphics[width=2.2in]{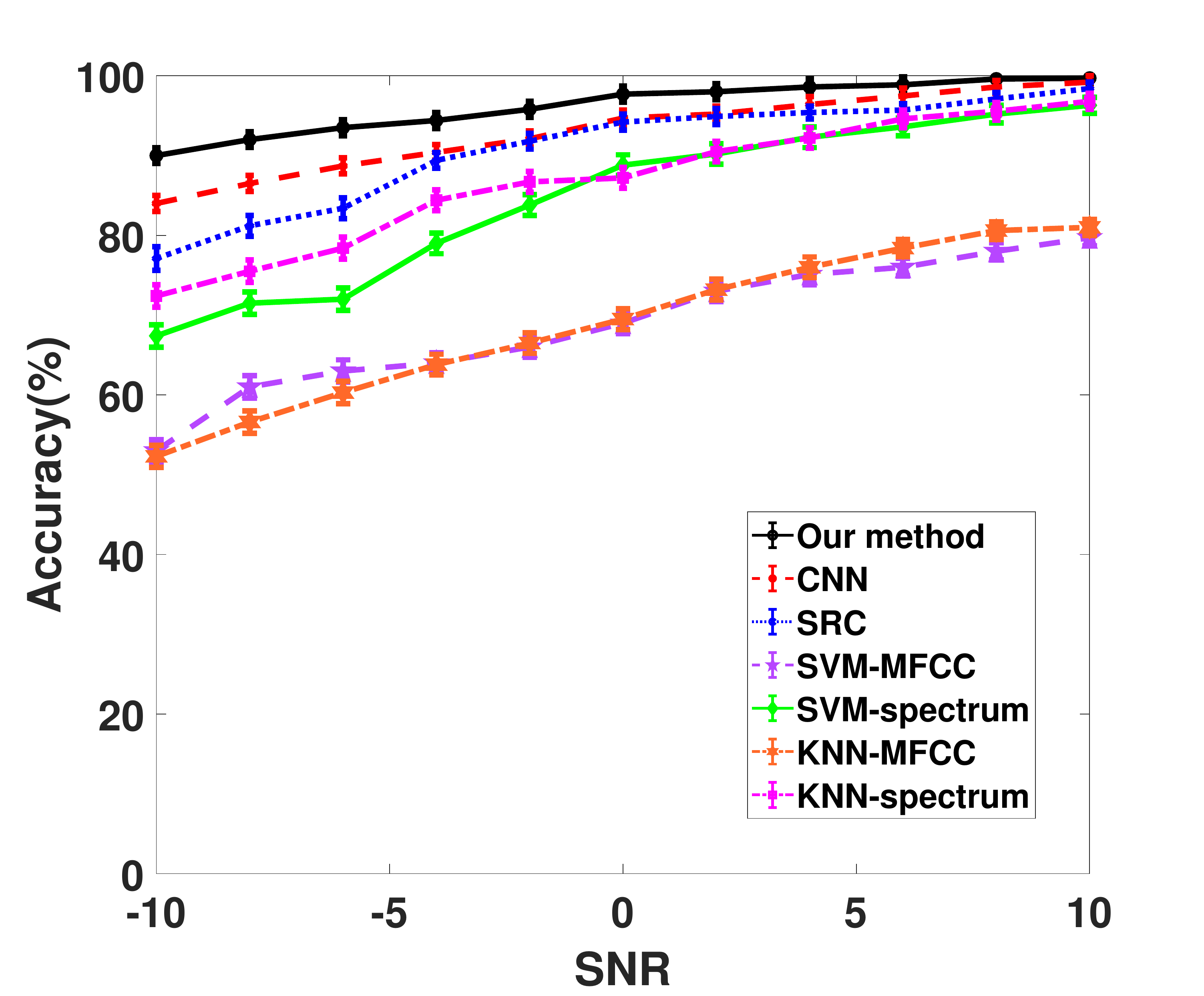}
        \label{fig:comparison_cricket2}
    } 
    \subfigure[Twenty-class classification.]{
         \includegraphics[width=2.2in]{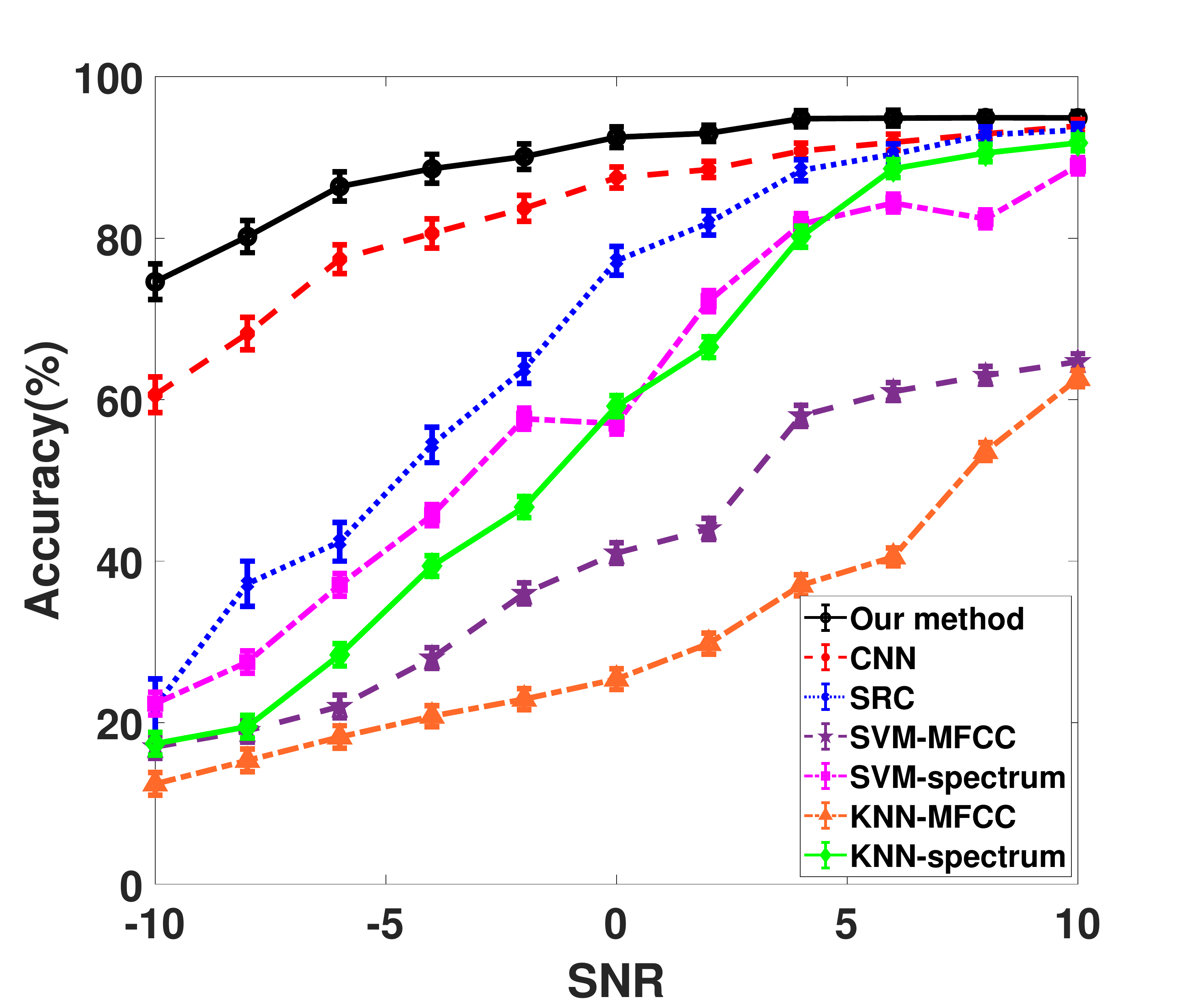}
        \label{fig:comparison_cricket20}
    }
    \caption{2 class classification vs 20 class classification.}
   \label{fig:cricketdataset}
\end{figure}

\subsection{Performance of Cricket Dataset}
\begin{figure*}[!ht]
    \centering
    \subfigure[Network topology]{
         \includegraphics[scale = 0.3]{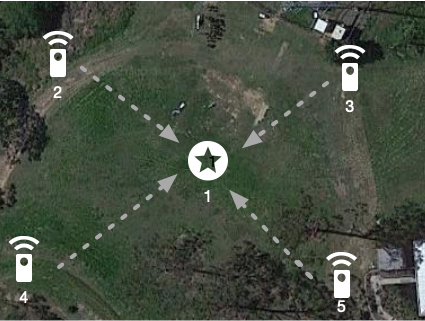}
        \label{fig:local}
    }
   \subfigure[Bird species]{
          \includegraphics[width=3in]{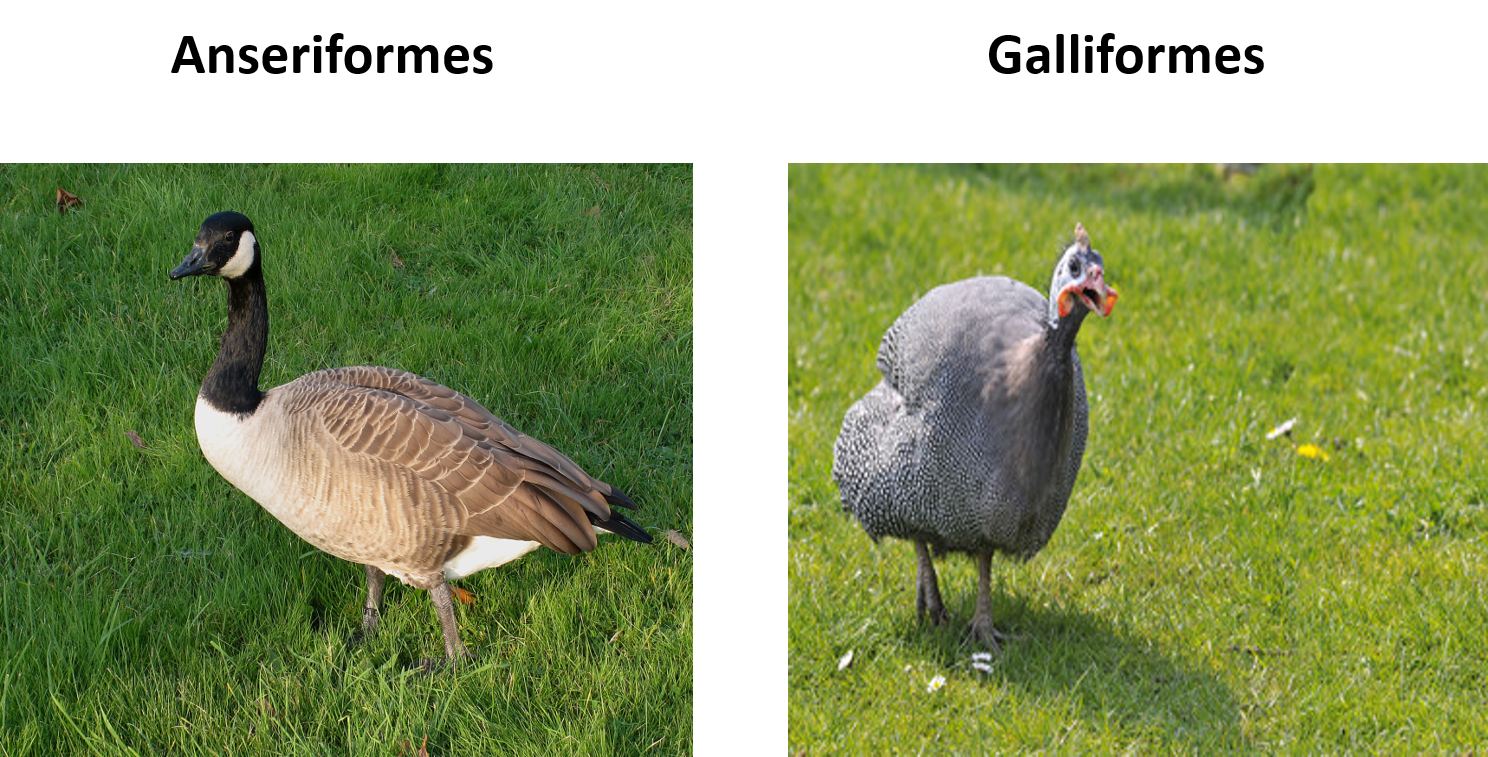}
       	 \label{fig:bird}
   }
    \caption{Testbed.}
  \label{fig:casestudy}
\end{figure*}

Similar to Frog dataset, we also evaluate the impact of window size, the number of iterations, the dropout rate and learning rate in CNN, and the size of training dataset using Cricket dataset. The procedures are the same as above and the results are shown in Figure~\ref{fig:cricketdataset}. We can see that it shows similar patterns as Frog dataset which suggests that the proposed framework is robust to different species. In terms of the dropout rate and learning rate, the optimal values for dropout rate and learning rate are 0.7 and 0.0005 which is slightly different from that of Frog dataset.

As mentioned in~\cite{hao2013monitoring}, the cricket dataset consists of twenty species of insects, eight of which are Gryllidae and twelve of which are Tettigoniidae. Thus, we can treat the problem as either a  two-class genus level problem, or twenty-class species level problem. We first treat the classification as a two-class level problem and calcualte the accuracy of different methods under different SNRs. From the results in Figure~\ref{fig:comparison_cricket2}, we can see that our method, SRC, SVM-spectrum and KNN-spectrum can achieve high accuracy. However, our method still outperforms all the other classifiers. Thereafter, we treat the classification as a twenty-class species level problem and plot the accuracy of different methods in Figure~\ref{fig:comparison_cricket2}. We can see that the proposed method significantly outperforms the other methods when SNR is low. Table~\ref{tab:comparison_cricket} summarizes the results of each method in detail. The results above demonstrate the advantage of our method in classifying more species in noisy environment.

\section{Case Study on Testbed}
\label{sec:testbed}
%To validate the feasibility of the proposed framework in real environment, we implement the system on an outdoor ASN testbed. The testbed is located on our campus and composed of five nodes which are configured as \emph{Ad-hoc} mode with a star network topology as can be seen in Figure~\ref{fig:local}. The goal of this case study is to evaluate the system's capability of recognizing bird vocalization in real world environment. 
%In this section, we show the feasibility of the proposed framework in real environment based on the testbed designed in~\cite{wei2013real}. 
% We implement the system based on the testbed designed in~\cite{wei2013real}. 
To validate the feasibility of the proposed framework in real environment, we implement the system on an outdoor ASN testbed which is located in Brisbane, Australia. As shown in Figure~\ref{fig:local}, the testbed is composed of five nodes which are configured as \emph{Ad-hoc} mode with a star network topology. Its task is to evaluate the system's capability of recognizing bird vocalization in real world environment. 

 \begin{table}[htb]
 \centering
     \caption{Power Consumption.}
     \label{tab:powerconsumption}
     \begin{tabular}{|l| c| }
     \hline
   \textbf{Module} & \textbf{Consumption (W)} \\
   \hline
     CPU  & 2.05\\
     \hline
     CPU + microphone & 2.1\\
     \hline
     CPU + Wifi (idle) & 2.45 \\
     \hline
     CPU + Wifi (Rx) & 2.67 \\
     \hline 
     CPU + Wifi (Tx) & 2.78 \\
         \hline
     \end{tabular}
     \label{tab:energy}
 \end{table}

The hardware platform used in the testbed is based on a Pandaboard ES with an 1.2Ghz OMAP 4460, 1GB Ram and 4GB SD-card. Additionally, Pandaboard includes an 802.11 interface for wireless connection. Microphones are connected to the Pandaboard via USB port to record bird voice with 24Khz sampling rate. All the nodes are connected via the local Wi-Fi network. The data collected from Node 2, 3, 4 and 5 will be first transferred to Node 1. Then, all the data will be uploaded from Node 1 to the local server. The acoustic date from different nodes are classified separately in the system.

In the testbed, each node is powered by a rechargeable battery (12V, 7.2Ah), and an optional solar panel (5W, 12V).  The power consumption of each module is given in Table~\ref{tab:powerconsumption}. Compared to SolarStore testbed~\cite{solarstore} which consumes 10W (low load) and 15W (high load) energy, our testbed is approximately 3.5 to 5.4 times more energy efficient. Without solar panel, a node in our ASN testbed will run continuously for more than 31 hours, which is significantly longer than the previous platforms such as ENSBox \cite{Girod2006:AENSBox}. We find that if a solar panel is exposed to direct sunlight for 8 hours per day, the node can maintain a 50\% duty cycle at 85\% solar charge efficiency.

The nodes use Network Time Protocol (NTP) for time synchronization. We use one node as the NTP server, and the other nodes as the NTP clients. The NTP clients send request for time synchronization every 10 seconds. The accuracy of time synchronization is about 25 ms, which is good enough for our distributed real-time system because the length of each testing signal segment is 400ms.

During deployment, we found that the recorded voice is deteriorated by wind. To solve this problem, we take two measures. First, we install foam and fur windscreen around each microphone. Second, we apply a Butterworth high pass filter with 200Hz cut-off frequency to filter out unwanted noise. This is because most of the wind audio energy lies in the frequency band below 200Hz, while most of the vocalization energy of the birds is in the frequency band higher than 200Hz.

After implementing the proposed framework on the testbed, we calculate the computation time on the node's side and classification accuracy on the server's side. On the node's side, we find that the node in our testbed can process all the captured acoustic data in real time. From Table~\ref{tab:performancetestbed}, we can see the silence removal and FFT take 20.38 ms and 15.33 ms, respectively.
 \begin{table}[]
 \centering
 \small
 \caption{Computation time of local processing.}
 \label{tab:performancetestbed}
 \begin{tabular}{ccc}
 \toprule
           & Silence Removal & FFT \\ \hline
 Time (ms) & 20.38 $\pm$ 2.04            & 15.33 $\pm$ 0.63             \\ \bottomrule
 \end{tabular}
 \end{table}
\begin{table*}[!h]
\centering
\small
\caption{Performance on testbed.}
\label{tab:performancetestbed}
%\resizebox{!}{0.38in}{
\begin{tabular}{cccccc}
\toprule
         & Our system          &CNN       & SRC   &SVM-Spectrum   & KNN-Spectrum\\ \hline
Accuracy & \textbf{90.3\%}   &84.4\%   &72.3\%   & 65.7\%       & 68.8\%  \\ \hline
Precision & \textbf{91.2\%}   &82.1\%   & 72.6\%  & 66.4\%       & 69.2\% \\ \hline
Recall    & \textbf{89.4\%}   &84.6\%   & 70.9\%  & 65.6\%       & 67.1\%  \\ \hline
F1-score  & \textbf{91.1\%}  &83.7\%   & 71.8\%  & 66.4\%       & 70.5\% \\ \bottomrule
\end{tabular}
%}
\end{table*}

In this study, we choose two common bird species in the area of interest: Anseriformes and Galliformes (Figure~\ref{fig:bird}). 
Our goal is to classify the voice into three classes: Anseriformes, Galliformes and others. The testbed runs for 30 days and the data is labeled manually. 
Table~\ref{tab:performancetestbed} lists the results of different methods for classification in the server. We find that the proposed system achieve $90.3\%$ classification accuracy which outperforms other methods significantly. The results in turn suggest that the proposed framework is robust to environmental noise and can achieve high classification accuracy in real-world WASN. We also notice that the results of the case study is slightly lower than the simulation results in Section~\ref{sec:evaluation}. This is because the public dataset are collected in a controlled manner and the signals are well trimmed and processed. However, the data we used in our case study are collected in a totally automatic manner.

\section{Conclusion}
\label{sec:conclusion}
In this paper, we design and implement a CNN-based acoustic classification system for WASN. To improve the accuracy in noisy environment, we propose a multi-view CNN framework which contains three convolution operation with three different filter length in parallel in order to extract the short-, middle-, and long-term information at the same time. Extensive evaluations on two real datasets show that the proposed system significantly outperforms previous methods. To demonstrate the performance of the proposed system in real world environment, we conduct a case study by implementing our system in a public testbed. The results show that our system works well and can achieve high accuracy in real deployments.  In our future work, we will deploy the proposed framework in wider area and evaluate its performance in different environments.

\section*{Acknowledgement}
The work described in this paper was fully supported by a grant from City University of Hong Kong (Project No.7200642)

\balance
\bibliography{main}

\end{document}